\documentclass[twocolumn]{article}

\usepackage[utf8]{inputenc}
\usepackage[top=2cm, bottom=3cm, left=2cm, right=2cm]{geometry}
\usepackage[T1]{fontenc}
\usepackage{authblk}
\usepackage{amsfonts}
\usepackage{amsmath}
\usepackage{amssymb}
\usepackage{xfrac}
\usepackage{graphicx}
\usepackage{float}
\usepackage{booktabs}
\usepackage{array}
\usepackage[backend=biber,style=phys]{biblatex}
\usepackage{comment}
\usepackage[skip=2pt,font=footnotesize]{caption}
\usepackage{titlesec}
\usepackage[flushleft]{threeparttable}
\usepackage{color}
\usepackage{xcolor}
\usepackage[normalem]{ulem}
\usepackage{tablefootnote}
\usepackage{caption}
\usepackage{subcaption}
\usepackage{hyperref}

\newcommand{\cmt}[1]{{\color{black!40!green}{}(#1)}}
\newcommand{\add}[1]{{\color{blue}{}#1}}
\newcommand{\del}[1]{{\color{red}{}\sout{#1}}}
\newcommand{\rpl}[2]{{\del{#1 }\add{#2}}}

\newcolumntype{L}[1]{>{\raggedright\arraybackslash}m{#1}}
\newcolumntype{M}[1]{>{\centering\arraybackslash}m{#1}}
\newcolumntype{R}[1]{>{\raggedleft\arraybackslash}m{#1}}

\titleformat*{\section}{\normalsize\bfseries\rmfamily}
\titleformat*{\subsection}{\normalsize\bfseries\rmfamily}
\titleformat*{\subsubsection}{\normalsize\bfseries\rmfamily}

\graphicspath{{./Figures/}}
\DeclareGraphicsExtensions{.jpg,.pdf,.pdf}
\title{\textbf{EuroPED-NN: Uncertainty aware surrogate model}}
\author[1,2,*]{A. Panera Alvarez}
\author[1,3]{A. Ho}
\author[4]{A. J{\"{a}}rvinen}
\author[5]{S. Saarelma}
\author[1,6]{S. Wiesen}
\author[a]{JET Contributors}
\author[b]{the ASDEX Upgrade Team}
\affil[1]{DIFFER -- Dutch Institute for Fundamental Energy Research, 5612 AJ Eindhoven, Netherlands}
\affil[2]{Eindhoven University of Technology, 5612 AZ Eindhoven, Netherlands}
\affil[3]{MIT Plasma Science and Fusion Center, Cambridge, 02139, MA, United States of America}
\affil[4]{VTT Technical Research Centre of Finland, FI-02044 VTT, Finland}
\affil[5]{United Kingdom Atomic Energy Authority, Culham Centre for Fusion Energy, Culham Science Centre, Abingdon, Oxon, OX14 3DB, UK}
\affil[6]{Forschungszentrum Jülich GmbH, Institut für Energie- und Klimaforschung - Plasmaphysik, DE-52425 Jülich,
Germany}
\affil[a]{See the author list of J. Mailloux et al. 2022 Nucl. Fusion 62 042026}
\affil[b]{See the author list of Zohm et al. 2024 Nucl. Fusion, U. Stroth et al. 2022 Nucl. Fusion 62 042006.}
\affil[*]{E-mail: \href{mailto:a.paneraalvarez@differ.nl}{a.paneraalvarez@differ.nl}}

\begin{document}
\twocolumn[
  \begin{@twocolumnfalse}
    \date{}
    \maketitle
    \begin{abstract}
	This work successfully generates an uncertainty-aware surrogate model of the EuroPED plasma pedestal model using the Bayesian neural network with noise contrastive prior (BNN-NCP) technique. This model is trained using data from the JET-ILW pedestal database and subsequent model evaluations, conforming to EuroPED-NN. The BNN-NCP technique has been proven to be a suitable method for generating uncertainty-aware surrogate models. It matches the output results of a regular neural network while providing confidence estimates for predictions as uncertainties. Additionally, it highlights out-of-distribution (OOD) regions using surrogate model uncertainties. This provides critical insights into model robustness and reliability. EuroPED-NN has been physically validated, first, analyzing electron density $n_e\!\left(\psi_{\text{pol}}=0.94\right)$ with respect to increasing plasma current, $I_p$, and second, validating the $\Delta-\beta_{p,ped}$ relation associated with the EuroPED model. This affirms the robustness of the underlying physics learned by the surrogate model. On top of that, the method was used to develop a EuroPED-like model fed with experimental data, i.e. an uncertainty aware experimental model, which is functional in JET database. Both models have been also tested in $\sim 50$ AUG shots.
    \end{abstract}
  \end{@twocolumnfalse}
]

\section{Introduction}
\label{sec:Introduction}

A tokamak fusion reactor must achieve a sufficiently high core plasma pressure, density and temperature in order initiate and sustain fusion reactions for net energy generation. The standard operational scenarios in tokamak devices as based on the high confinement mode (\emph{H-mode})~\cite{Wagner_2007}. This regime is characterized by the formation of a transport barrier in the outer region of the confined plasma, leading to steep gradients in the density and temperature profiles within that region. As the core plasma parameters are effectively an integration of the transport equation from the edge boundary condition, this steep gradient region lifts up the core parameters and gives this feature its namesake, the plasma \emph{pedestal}.

This edge transport barrier has been linked to an increase in the radial electric field, $E_r$, shear~\cite{aRadialElectricField-Groebner} which drives a local bulk rotation shear that breaks up large turbulent eddies~\cite{aRotationShearing-Burrell}, suppressing the dominant transport mechanisms observed in that region in L-mode regimes~\cite{Diamond_2005}. It should be noted that the steep gradient region still contains turbulent transport processes, although they are driven by other mechanisms which are less affected by the flow shear~\cite{Curie_2022,Ren_2022,Hatch_2021}. While the precise mechanism of the $E_r$-well formation for the L-H transition is still under investigation~\cite{Solano_2022,Vermare_2022,Silva_2021}, the resulting transport reduction causes the plasma gradients to increase, along with the plasma current due to the bootstrap effect~\cite{Peeters_2000}, up until a point where MHD instabilities occur, e.g. edge-localized modes (ELMs)~\cite{aEdgeLocalizedModes-Leonard}. These ELMs manifest as a discrete and periodic collapse of the pedestal, effectively reinstating another, albeit higher, limit on the pressure gradient. The predominant explanation of ELMs goes via MHD theory, specifically the interaction between instabilities in the pressure driven (ballooning) modes and current driven (peeling) modes~\cite{aPeelingBallooning-Wilson}, due to these steep gradients in the edge region. Within the EPED model~\cite{EPED_paper}, this theoretical pedestal limit is then located where the pedestal pressure, $p_{ped}$, and width, $\Delta$, lie on the intersection of the onset of ideal MHD instabilities and the onset of kinetic ballooning mode (KBM) turbulence, which the latter is represented by a semi-empirical relation. The large type-I ELMs are well described by this model~\cite{Zohm_1996}, tracing out stability and instability regions in the current density-pressure gradient ($j$-$\alpha$) space. The EuroPED model~\cite{aJETEuropedDatabase-Saarelma} effectively predicts these pedestal limits along these same theoretical principles.

Unfortunately, the computational cost of both EuroPED and EPED is too high for highly iterative applications such as plasma scenario optimization~\cite{aRAPTORQLKNN-VanMulders} and motivates the development of a surrogate model which uses less computational resources while to representing the original model reasonably accurately. Various approaches in the fusion community utilize machine learning algorithms to develop surrogate models and their subsequent applications~\cite{aMLShapeControl-Degrave,aQLKNN-Ho,aQLKNN-vdPlassche,aTGLFEPEDNN-Meneghini,aMMMNet-Morosohk,aDisruptionHybridML-Zhu,aNNDisruption-KatesHarbeck,Meneghini_2021,Zeger2021}. The fast computational throughput also opens the door to explore the behaviour of the model and the physics behind it. This work investigates the applicability of Bayesian neural networks (BNNs) to generate a pedestal surrogate model, potentially incorporating their ability to provide uncertainty information in a single forward pass as part of a prediction application. This approach aligns with previous work in using machine learning to enhance plasma pedestal modeling \cite{Gillgren_2022, aTGLFEPEDNN-Meneghini}.

In the Bayesian approach, the variables in the neural networks are treated as distributions, which are propagated throughout the calculation to provide statistical information about the model prediction, labelled in this study as output \emph{uncertainties}. The training process needs to determine the distributions of each output using a variational inference approximation~\cite{aVariationalBayes-Kingma} allowing the construction of a feasible loss function. However, the main drawback of variational inference is the high degree of sampling necessary for the training. This is circumvented using another acceleration technique called the Noise Contrastive Prior (NCP)~\cite{aBNN_NCP-Hafner}, which has a secondary advantage of estimating both the \emph{epistemic} and \emph{aleatoric} uncertainties. The epistemic uncertainties (or model uncertainties) are defined as uncertainties due to the lack of knowledge about the system, such as missing input parameters or insufficent data volume, and the aleatoric uncertainties (or data uncertainties) are defined as uncertainties resulting from random chance, such as repeat measurements or experiments providing different results~\cite{Hullermeier_2021,Kiureghian_2009}. This study expands on the methodology and interprets the results of the trained model based on these definitions.

Section~\ref{sec:Methodology} describes the specific application of the BNN-NCP model in this study, including the considerations in determining the user-defined parameters impacting the aleatoric and epistemic uncertainty predictions. In Section~\ref{sec:Verification}, the precision and practicality of this model is discussed, focusing on its capacity to accelerate computations, gather physics insights, and adapt to a different machine. Section \ref{sec:exp_model} extends the methodology developed to experimental JET-ILW data and tests it in AUG data. Finally, Section~\ref{sec:Conclusion} provides a summary of the main conclusions and potential future work inspired by this study.

\section{Development of EuroPED-NN}
\label{sec:Methodology}

In this section, the implementation of the BNN-NCP model towards building a plasma pedestal surrogate model is detailed.

\subsection{Model description, inputs and outputs}
\label{subsec:ModelInputOutput}

The EuroPED model is conceptually identical to the EPED model, which uses an ideal MHD stability description to predict the critical plasma pressure gradient, $\alpha$, as a function of the pedestal width, $\Delta$. This curve is then crossed against a semi-empirical $\Delta$-$\beta_{p,ped}$ relation reflecting the KBM constraint~\cite{EPED_paper} to provide a prediction of the pedestal properties. The assumption of a two-species plasma in equilibrium is then used to estimate the top-pedestal electron pressure, $p_{e,ped}$, and further decompose that into the top-pedestal electron temperature, $T_{e,ped}$, by requiring the top-pedestal electron density, $n_{e,ped}$, as input.

Thus, for the EPED model, the required input parameters include:
\begin{itemize}
    \itemsep 0mm
    \item plasma parameters:
    \begin{itemize}
        \itemsep 0mm
        \item plasma current, $I_p$, in [MA];
        \item toroidal magnetic field, $B_t$, in [T];
        \item line-integrated effective charge, $Z_{\text{eff}}$;
        \item top-pedestal electron density, $n_{e,ped}$ , in [$\times10^{19}$ m\textsuperscript{-3}];
        \item normalized plasma pressure, $\beta_N$;
    \end{itemize}
    \item and magnetic geometry parameters:
    \begin{itemize}
        \itemsep 0mm
        \item triangularity, $\delta$;
        \item elongation, $\kappa$;
        \item minor radius, $a$, in [m];
        \item major radius, $R_0$, in [m].
    \end{itemize}
\end{itemize}
In practice, one drawback of the EPED model is the necessity of providing $n_{e,ped}$, $\beta_N$ and $Z_{eff}$ as inputs, as this precludes the model from predictive exercises as these parameters are often unknown before performing an actual experiment.

To remedy this, the EuroPED model was developed by coupling an EPED-like pedestal parameter calculation workflow to the Bohm/gyro-Bohm (BgB) turbulent transport model~\cite{Erba_1998} inside a 1D transport solver. This removes the need to explicitly define $\beta_N$ and $n_{e,ped}$, as the former is provided by the kinetic profiles resulting from the self-consistent transport calculation and the latter is defined either via a regression or a neutral penetration model provided within EuroPED. However, this model additionally requires two engineering parameters as input:
\begin{itemize}
    \itemsep 0mm
    \item injected auxiliary heating power, $P_{tot}$;
    \item separatrix electron density, $n_{e,sep}$.
\end{itemize}
This is seen as an improvement over the EPED model as $\beta_N$ or $n_{e,ped}$ are difficult to estimate prior to conducting an experiment. Although often the experiments are run in feedback mode where these parameters can actually be chosen. However, from the model point-of-view EPED is still not fully predictive. While the $n_{e,sep}$ is equally difficult to know prior to an experiment, the option to estimate it via the fuelling gas flow rate, $R_{\text{gas}}$, and a neutral penetration model is provided within EuroPED. This connection was excluded from this study as it was not used in the construction of the particular database forming the training set.

The EuroPED model then returns as outputs:
\begin{itemize}
    \itemsep 0mm
    \item the top-pedestal electron density, $n_{e,ped}$;
    \item the top-pedestal electron temperature, $T_{e,ped}$;
    \item and pedestal width, $\Delta$.
\end{itemize}
However, due to the implementation details connecting the EPED-like workflow to the transport model, the actual radial locations of the top-pedestal quantities may not coincide in practice. This has some implications on the construction of an appropriate surrogate model, which is further discussed in Section~\ref{subsec:Dataset}.


\subsection{BNN-NCP Technique}
\label{subsec:BNN-NCP}

As mentioned earlier, Bayesian neural networks are a specific class of NNs trained using Bayesian inference methods, involving an estimation of a posterior distribution over network weights given observed data. One of the difficulties with training BNNs is that they can be computationally expensive, particularly for large networks with many parameters. By providing a fast and efficient approximation to the posterior distribution, the Noise Contrastive Prior technique can speed up BNN training.

Within this technique, the posterior distribution is approximated by a mixture of the prior distribution over the output and a noise distribution. The network is trained to distinguish between observed data and noise samples generated by the noise distribution during training. Essentially, it uses simulated noise to shift the hypothesis from the probability distribution of the neural network weights to a probability distribution in the data space, requiring less computational resources. In practice, this means that a random point from the joint input distribution is sampled per training epoch. This sampled point is passed forward through the current BNN configuration, and the resulting mean and epistemic uncertainty is compared to a user-defined prior output distribution via some distance metric. This significantly reduces the number of samples required per epoch compared to a typical Monte Carlo method by leveraging the large number of epochs required for training the model to effectively accumulate the necessary statistics, which is similar to the training advantages offered by mini-batching and stochastic gradient descent. Furthermore, the NCP method improves the detection of out-of-distribution (OOD) regions in data space by directly accounting for input variations within its epistemic uncertainty estimation. In summary, this method allows for an approximate calculation of the output distribution without the need to sample the weights to arrive at a reliable estimate of the uncertainty. This approach is presented in \cite{aBNN_NCP-Hafner} for one input and one output problems, in this study a multidimensional approach to fit the needs of the plasma pedestal models is built.

\begin{figure*}[tb]
    \centering
    \begin{subcaptionblock}{10cm}
        \centering
        \includegraphics[width=0.7\linewidth]{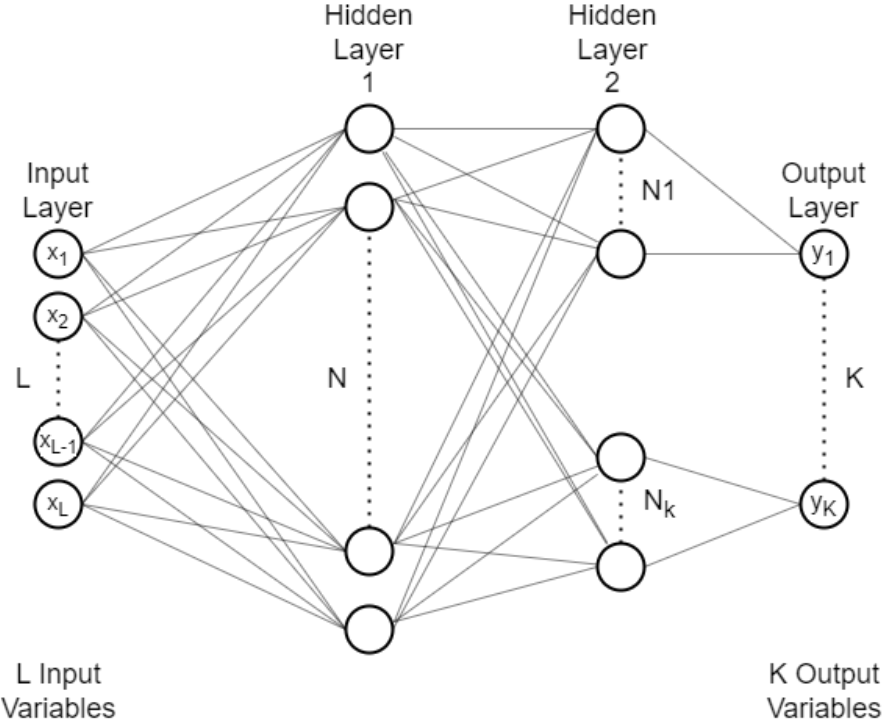}
        \caption{Main network scheme}
        \label{fig:arch_1}
    \end{subcaptionblock}%
    \begin{subcaptionblock}{4cm}
        \centering
        \includegraphics[width=\linewidth]{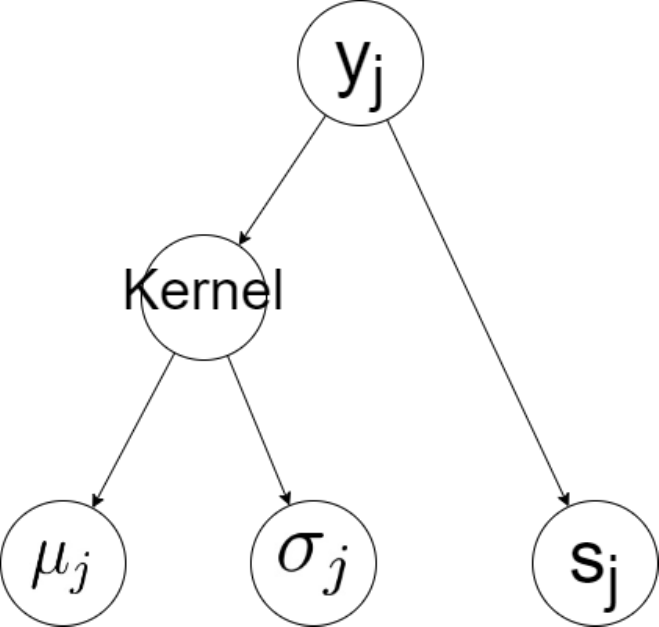}
        \caption{Output layer scheme}
        \label{fig:arch_2}
    \end{subcaptionblock}
    \caption{Architecture schema of the BNN-NCP developed in this study. In (a), L and K represents the number of input and output variables respectively; N, the number of neurons in the first layer; $N_1\:...\: N_K$ represent the number of neurons in each of the K packs of neurons associated with every output neuron. In (b), $y_j$ represents a output neuron; kernel, the probability distribution of weights and bias; $\mu_j$ and $\sigma_j$, the mean and standard deviation of the Kernel; and $s_j$, an independent neuron.}
\end{figure*}

\subsection{Neural network architecture and loss function}
\label{subsec:LossFunction}

As a brief overview, the BNN-NCP architecture consists of a number of fully connected feed-forward (i.e. dense) layers with a custom layer inserted just before the output layer, as shown in Figure~\ref{fig:arch_1}. This custom layer consists of a \emph{variational layer} and a dense layer with a ``softplus" activation function, as shown in Figure~\ref{fig:arch_2}, each with only a single neuron. This variational layer represents its weights and biases as joint probability distribution, also known as a \emph{kernel}, which effectively allows them to be sampled to assemble its predictive distribution, $p\!\left(\mathbf{y}^*|\mathbf{x}^*,\mathbf{x},\mathbf{y},\theta\right)$, where the $^*$ notation is used to differentiate a generic point from one which is explicitly within the training dataset. The ``softplus" activation function on the dense layer enforces its output to be positive, allowing it to represent an independent uncertainty estimate. All other neurons in the network have a ``leaky ReLU" activation function, which is effectively a piecewise function of two linear components, both with non-zero gradients and joined together at the origin.

Due to the extension of this technique to multiple output variables, the final dense layer before the custom layer is separated into dedicated blocks per output variable. This serves as an attempt to ease the training process by reducing the amount of cross-talk between the network predictions of each variable. It also introduces a degree of flexibility in the last hidden layer, allowing adjustments to suit the complexity of each variable, e.g. the most difficult variables to fit can have a large number of neurons to account for their dependencies. This results in a multi-input multi-output model, transforming input points, $\mathbf{x} = \left\lbrace x_i \right\rbrace$, to output points, $\mathbf{y} = \left\lbrace y_j \right\rbrace$, which can then be trained on a labelled training dataset, denoted as $D = \left\lbrace \left(\mathbf{x}, \mathbf{y}\right)_k \right\rbrace$.

As a result of the BNN-NCP architecture, there are two modes of evaluating it on a given input point, $\mathbf{x}$:
\begin{enumerate}
    \itemsep 2mm
    \item By sampling the probability distribution represented by the kernel in the variational layer, resulting in a stochastic output, $\mathbf{\hat{y}}$;
    \item By computing the joint statistical moments of the kernel in the variational layer, resulting in a deterministic mean output, $\mathbf{\mu}$, and its standard deviation, $\mathbf{\sigma}$.
\end{enumerate}
Both of these evaluation modes return another deterministic output from the softplus neuron, $\mathbf{s}$. The predictive output value of the BNN-NCP model is taken as the mean value, i.e. $\mathbf{y}^* \equiv \mathbf{\mu}$, from the second evaluation method due to its deterministic nature, although the first evaluation method is still described here due to its importance for the training methodology. Thus, the entire BNN-NCP model can be mathematically represented as a black box function, $\left\lbrace \mathbf{\mu}, \mathbf{\sigma}, \mathbf{s} \right\rbrace^* = f\!\left(\mathbf{x}^*\right)$.

Then, based on these descriptions, the output standard deviation from taking the statistical moment of the kernel, $\mathbf{\sigma}$, was labelled as the epistemic uncertainty prediction and the output of the independent softplus neuron, $\mathbf{s}$, as the aleatoric uncertainty prediction. Both uncertainty outputs are taken to represent a value of one standard deviation when converting them from statistical moments back into probability distributions.

The cost function used in this study is similar to the one introduced in Ref.~\cite{aBNN_NCP-Hafner} except extended into the multidimensional approach. This results in expression provided in Equation~\eqref{eq:loss_BNN-NCP_multi}. The terms are then summed over each output variable, represented by the subscript $j$.
\begin{equation}
    \begin{aligned}
    \mathcal{L}\!\left(\theta\right) = &-\sum_j \gamma_{\text{nll},j} \cdot \ln{p\left(y_j|\mathbf{x},\theta\right)}\\
    &+ \sum_j \gamma_{\text{epi},j} \cdot \text{KL}\left(p_{\text{prior,epi},j} \, || \, q\!\left(\tilde{\mu}_j|\mathbf{\tilde{x}},\theta\right)\right)\\
    &+ \sum_j \gamma_{\text{alea},j} \cdot \text{KL}\left(p_{\text{prior,alea},j} \, || \, \mathcal{N}\!\left(0,\tilde{s}_j^2\right)\right)
    \end{aligned}
    \label{eq:loss_BNN-NCP_multi}
\end{equation}
where $\gamma$ represents a user-defined weight coefficient used to adjust the associated loss term, $\theta$ represents the distribution of weights in the BNN, and the $\sim$ notation is used to indicate quantities related to the OOD region sampling. The sampled OOD input values, $\mathbf{\tilde{x}}$, were generated from a normal distribution about each dataset point, $\mathbf{x}$, according to a user-defined input noise term, $\mathbf{\Sigma_x}$. This input noise should generally be chosen as a large enough value to ensure the OOD sampling goes adequately beyond the training dataset input values, $\mathbf{x}$, but not so large that the OOD sampling would have difficulty resolving well the area just outside the training dataset.

The first term of the loss function corresponds to the negative log-likelihood (NLL) which represents a goodness-of-fit metric for the approximate predicted posterior distribution, selected to be:
\begin{equation}
    p(y_j|\mathbf{x},\theta) \sim \mathcal{N}\!\left(\hat{y}_j,s_j\right)
    \label{eq:nll_posterior}
\end{equation}
By minimizing this term with respect to the weights and biases, $\theta$, the maximum likelihood value is obtained for the network prediction. As the aleatoric uncertainty prediction, $s_j$, is used as the weighting parameter for this metric, it also encourages the aleatoric uncertainty to grow where the network cannot match the specific data point where the combination of other data points and regularization discourage it, e.g. where data noise is present.

The second term represents a metric to anchor and regularize the predicted epistemic uncertainty with weight coefficients, $\gamma_{\text{epi},j}$, which are used to adjust its relative importance in the loss function per output variable. The last term represents the same anchoring and regularization metric for the predicted aleatoric uncertainty, with its own relative weight coefficients per output variable, $\gamma_{\text{alea},j}$.

Both uncertainty regularization terms use the Kullback-Leibler (KL) divergence metric, which effectively compares the similarity of two distributions. For the epistemic term, the approximate output distribution predicted by the network using an OOD input distribution, $q\!\left(\mu_j|\mathbf{\tilde{x}},\theta\right)$, was compared against an epistemic prior distribution, $p_{\text{prior,epi},j}$, selected to be:
\begin{equation}
    p_{\text{prior,epi},j} = \mathcal{N}(y_j, \sigma_{y_j}^2)
    \label{eq:kl_prior_epi}
\end{equation}
This represents a normal distribution where the mean is the corresponding output value in the dataset, $y_j$, and the standard deviation, $\sigma_{y_j}$, is a user-defined parameter encapsulating any expected epistemic uncertainty already known to be associated with the data-generating system. This term is expected to generate a significant loss for OOD samples far away from the training point, effectively encouraging the epistemic uncertainty at those OOD samples to grow in order to minimize the overall loss. In order to save computational time during training, only one OOD point is sampled per training point per epoch, which relies on the stochastic gradient descent process to generalize this feature to the entire OOD domain and is one of the essential acceleration tricks provided by the NCP methodology. For the aleatoric term, a normal distribution with zero mean created from the aleatoric uncertainty prediction of the network, $\mathcal{N}(0,s_j^2)$, was compared against an aleatoric prior distribution, selected to be:
\begin{equation}
    p_{\text{prior,alea},j} = \mathcal{N}(0,s_{y_j}^2)
    \label{eq:kl_prior_alea}
\end{equation}
Again, this represents another normal distribution where the standard deviation, $s_{y_j}$, is another user-defined parameter encapsulating any expected aleatoric uncertainty already known to be associated with the dataset output values. This is effectively meant to provide a competing pressure on the aleatoric uncertainty prediction, $s_j$, preventing it from growing to the point where a flat prediction with infinite aleatoric uncertainties becomes the most likely solution.




Basically, this way of treating the output results into the loss function is the main difference with respect to a regular neural network. For this matter, it is particularly meaningful to define several parameters that will help to adjust for the problem to solve. These user-defined values, $\mathbf{\Sigma_x} = \left\lbrace \sigma_{x_i} \right\rbrace$, $\mathbf{\Sigma_y} = \left\lbrace \sigma_{y_j} \right\rbrace$ and $\mathbf{s_y} = \left\lbrace s_{y_j} \right\rbrace$, are meaningful for the model as they are hyperparameters that introduce knowledge from the system and the data region to be modelled into the neural network. While in principle, this would allow extensions to the labelled training dataset, $D' = \left\lbrace \left(\mathbf{x}, \mathbf{y}, \mathbf{\Sigma_x}, \mathbf{\Sigma_y}, \mathbf{s_y}\right)_k \right\rbrace$, for refining the uncertainty predictions, it was decided to leave these extra hyperparameters constant across the entire original training dataset, $D$, for this study. This is primarily for ease of demonstration for the BNN-NCP methodology, as it simultaneously reduces the burden of collecting the information necessary to refine these priors and the complexity needed to be captured by the training process. More details about the values used for these priors is available in Section~\ref{subsec:HyperparameterTuning}.


\subsection{Zero-Dimensional Pedestal Database}
\label{subsec:Dataset}

During this study, the dataset is derived from a specialized collection JET-ILW plasma discharges~\cite{aJETExperimentalPedestalDatabase-Frassinetti} and their corresponding EuroPED predictions~\cite{aJETEuropedDatabase-Saarelma}. While the dataset includes several plasma quantities, this study focused on the input and output variables needed to predict the appropriate pedestal characteristics within a given plasma scenario accroding to EuroPED, as discussed in Section~\ref{subsec:ModelInputOutput}.

However, in order to improve the generality of the model via the introduction of dimensionless variables, the inverse aspect ratio, $\epsilon=a/R_0$, was used instead of the minor radius, $a$, and major radius, $R_0$. This process was also applied to the plasma current, $I_p$, replacing it by a safety-factor-like parameter, $\mu$, expressed as:
\begin{equation}
    \mu = \frac{\mu_0}{2 \pi a} \frac{I_p}{B_t}
    \label{eq:mu}
\end{equation}
where $I_p$ is expressed in units of A, $B_t$ in T and $a$ in m. While it is uncertain whether pedestal phenomena follow any specific dimensionless scaling, it was attempted to improve its extrapolation capability to other tokamaks.


However, due to the particular application of the pedestal predictions inside the 1D transport solver within EuroPED, the exact top-pedestal temperature and density outputs of the ideal MHD calculation are no longer representative of the plasma state described by the profiles. Thus, the variables provided in this dataset used as the target outputs were:
\begin{itemize}
    \itemsep 0mm
    \item electron density at $\psi_{\text{pol}} = \psi_0$, $n_e\!\left(\psi_0\right)$;
    \item electron temperature at $\psi_{\text{pol}} = \psi_0$, $T_e\!\left(\psi_0\right)$;
    \item and pedestal width, $\Delta$
\end{itemize}
where $\psi_{\text{pol}}$ is the poloidal magnetic flux, which is used as a geometry-independent radial coordinate, and $\psi_0 \equiv 0.94$ for the purposes of this study. This specific flux coordinate value was chosen due to its statistical similarity to the $\beta_{\text{pol},ped}$ computed from EuroPED.

The pedestal dataset is composed of 1317 entries. However, 257 of them were removed in the training and testing because some of the variables of interest were not available. During the training of the model, the dataset was split into 90\% for training and 10\% for testing.

\begin{figure*}
    \centering
    \includegraphics[width=\linewidth]{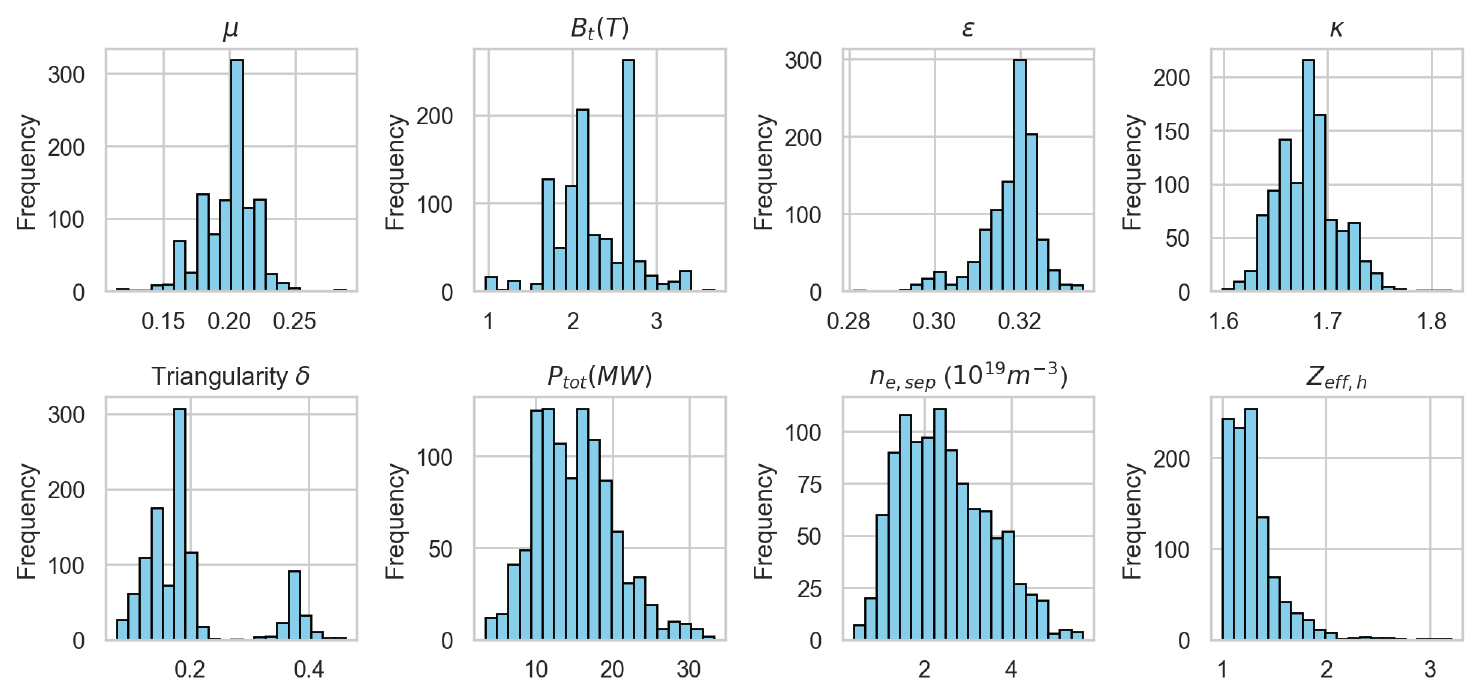}
    \caption{Input data histograms for EuroPED-NN from the dataset used in this study, obtained from~\cite{aJETExperimentalPedestalDatabase-Frassinetti}. $\mu$ stands for the parameter defined in equation \ref{eq:mu}, $B_t$ for toroidal magnetic field, $\epsilon$ for inverse aspect ratio, $\kappa$ for elongation parameter, $\delta$ for triangularity, $P_{tot}$ for injected auxiliary heating power, $n_{e,sep}$ for separatrix electron density and $Z_{eff,h}$ for line-integrated effective charge.}
    \label{fig:hist}
\end{figure*}

In Figure~\ref{fig:hist}, the input data (train and test) used in EuroPED-NN is shown as histograms. The distribution of all the input variables as well as the range of values can be observed.

\subsection{Hyperparameters for epistemic and aleatoric uncertainty}
\label{subsec:HyperparameterTuning}

As observed in the loss function, Equation~\eqref{eq:loss_BNN-NCP_multi}, several user-defined parameters and hyperparameters related with both the epistemic and aleatoric uncertainty predictions need to be determined. These values to the training routine such that the method to the particular problem. Therefore, additional considerations were required to provide reasonable targets for these values in training the EuroPED-NN.

The width of the OOD region for input sampling, $\mathbf{\Sigma_x}$, was chosen to span a constant proportion of the standard deviation within the dataset of the corresponding input quantity. These specific proportionality values are detailed Appendix~\ref{sec:tables}.


Regarding the target epistemic uncertainty, $\mathbf{\Sigma_y}$ in Equation~\eqref{eq:kl_prior_epi}, a small constant value was chosen across the entire dataset. So that, when comparing, using KL divergence, in the training loop the target distribution and the output distribution with OOD input, the OOD output distribution will be forced to become wider to match the thin target distribution.

Regarding the target aleatoric uncertainty, $\mathbf{s_y}$ in Equation~\eqref{eq:kl_prior_alea}, another small constant value was chosen across the entire dataset. This represents the prior knowledge that the EuroPED model is deterministic and thus, no data error should be present within the output values of the training dataset. The specific values used in this study are available in Appendix~\ref{sec:tables}.

All of these selections have been the result of different attempts and the personal reasoning of the authors regarding the nature of the model. Further details of these sensitivities are provided in Section~\ref{sec:Verification}, but the main reason is due to the fact that the epistemic and aleatoric uncertainties has been designated their own loss terms, each with their own weight coefficients, $\gamma_{epi}$ and $\gamma_{alea}$, in Equation~\eqref{eq:loss_BNN-NCP_multi}.

 
Once these parameters informing the BNN-NCP priors have been decided, a rudimentary configuration optimization exercise yielded an architecture with a common dense layer with 20 neurons and specialized dense layers with 8, 8, and 10 neurons for $n_e\!\left(\psi_0\right)$, $T_e\!\left(\psi_0\right)$, and $\Delta$ outputs, respectively. Then, the BNN was trained using the \textit{Keras} package \cite{chollet2015keras} in \textit{Tensorflow} \cite{tensorflow2015-whitepaper} (version 2.15.0) with 2000 training epochs and a mini-batch size of 53, taking around 1.7~CPU-hours (10~minutes in 10~cores $\approx 100$~minutes). The \textit{Adam} optimizer \cite{DBLP:journals/corr/KingmaB14} was used with a learning rate of 0.001 and a exponential decay rate of 0.1.

In order to balance the relative importance of the various loss terms in Equation~\eqref{eq:loss_BNN-NCP_multi}, a rudimentary optimization was performed. This resulted in the uncertainty weight coefficients of $\gamma_{\text{epi}} = 0.1$ and $\gamma_{\text{alea}} = 0.1$ for all outputs, $j$, and individual NLL weight coefficients of $\gamma_{\text{nll}} = \left\lbrace 7, 7, 2 \right\rbrace$ for $\Delta$, $T_e\!\left(\psi_0\right)$, and $n_e\!\left(\psi_0\right)$, respectively.

    


\section{Verification and Characterization of EuroPED-NN}
\label{sec:Verification}

Using the methods and principles described previously, the surrogate model of EuroPED, referred to as EuroPED-NN within this study, was trained. This section details the various methods used to verify its performance via whether it replicates expected behaviours.

\subsection{Performance}
\label{subsec:Performance}

\begin{figure*}[tb]
    \centering
    \begin{subcaptionblock}{6cm}
        \centering
        \includegraphics[width=\linewidth]{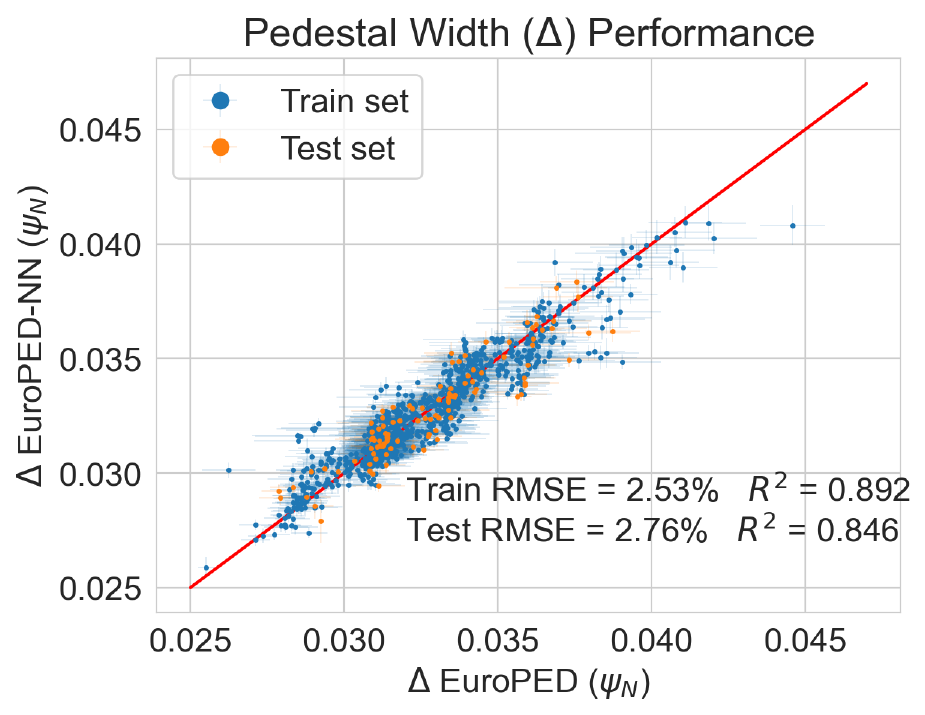}
        \caption{$\Delta$}
        \label{fig:delta_perf}
    \end{subcaptionblock}%
    \begin{subcaptionblock}{6cm}
        \centering
        \includegraphics[width=\linewidth]{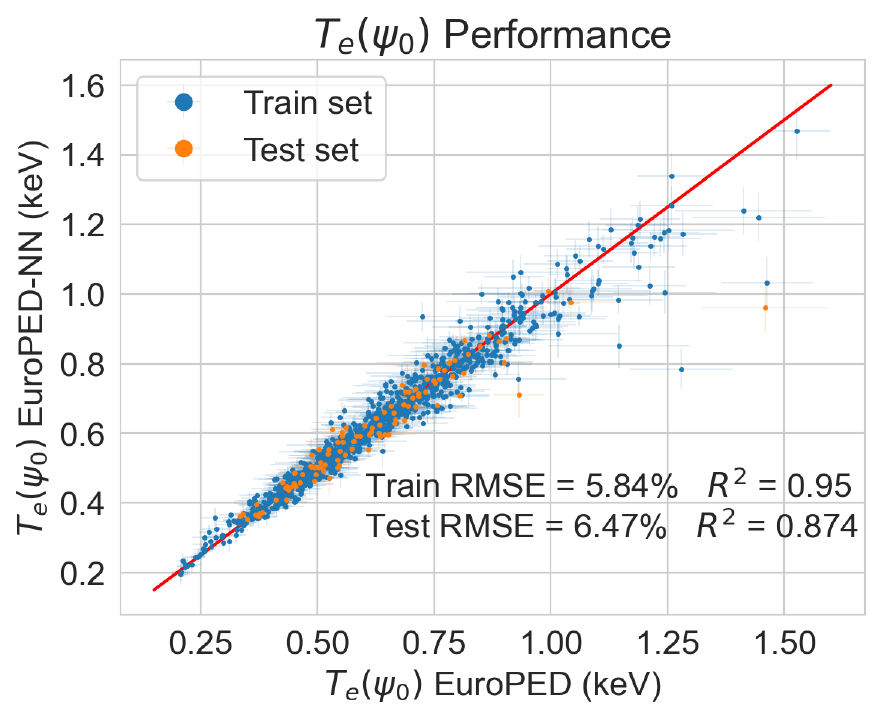}
        \caption{$n_{e,ped}$}
        \label{fig:teped_perf}
    \end{subcaptionblock}%
    \begin{subcaptionblock}{6cm}
        \centering
        \includegraphics[width=\linewidth]{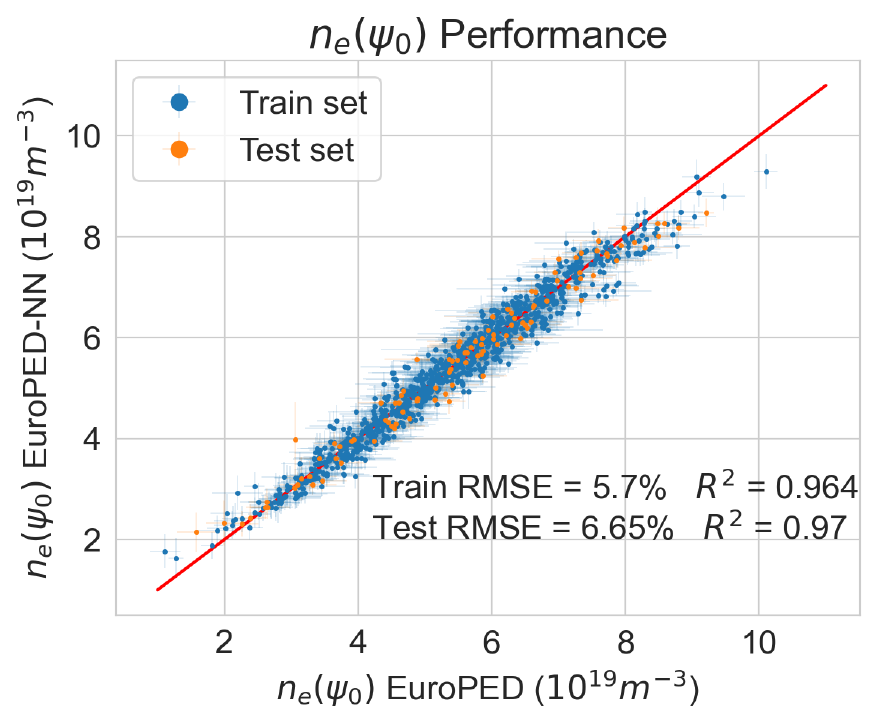}
        \caption{$T_{e,ped}$}
        \label{fig:neped_perf}
    \end{subcaptionblock}
    \caption{EuroPED-NN performance for its three output variables. Train and test data are shown in blue and orange respectively.  Epistemic and aleatoric uncertainties are displayed in vertical and horizontal axis respectively.}
    \label{fig:perf}
\end{figure*}

To test the performance of the model, the EuroPED-NN predictions are first compared against the data points within the training set itself. Figure~\ref{fig:perf} shows the accuracy for the output variables pedestal width, $\Delta$, electron temperature, $T_{e}\!\left(\psi_0\right)$, and electron density, $n_{e}\!\left(\psi_0\right)$, respectively. While they are all reasonably accurate, it can be visually inferred that $\Delta$ returns a worse match with the EuroPED database. This may indicate a higher degree of complexity behind the determination of this particular output, as it was found that it already required more neurons in the last BNN layer to achieve this fit quality. Overall, it can be said that the accuracy obtained through the training of the neural network is remarkably good, in spite of the typical difficulties of training multi-output regression networks.

\begin{figure*}[!ht]
    \centering    
    \begin{subcaptionblock}{8cm}
        \centering
        \includegraphics[width=0.9\linewidth]{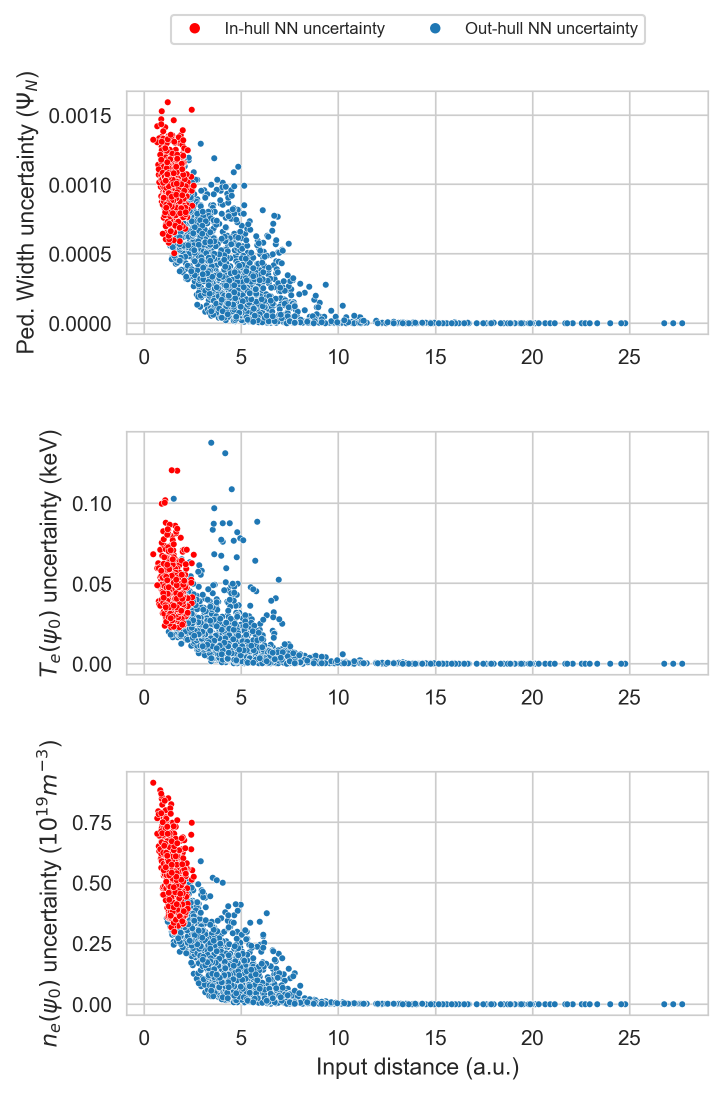}
        \caption{Absolute epistemic uncertainties}
        \label{fig:rel_alea}
    \end{subcaptionblock}%
    \begin{subcaptionblock}{8cm}
        \centering
        \includegraphics[width=0.9\linewidth]{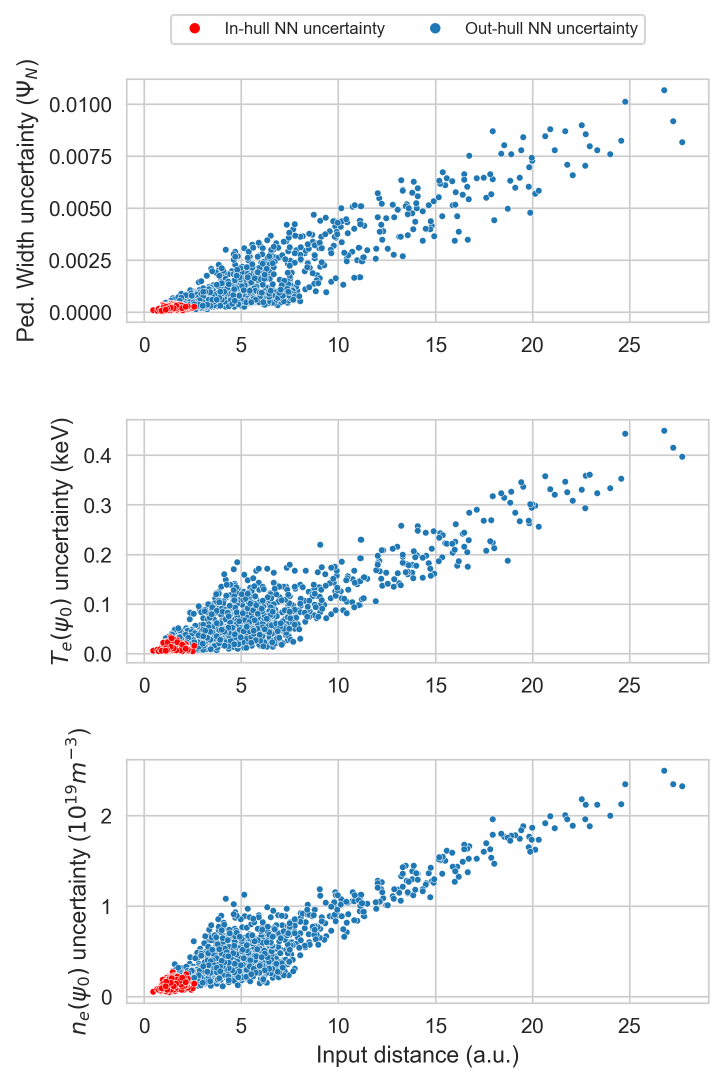}
        \caption{Absolute aleatoric uncertainties}
        \label{fig:rel_epi}
    \end{subcaptionblock}
    \caption{Absolute uncertainties of the EuroPED-NN surrogate model predictions against the distance from the input position to the input centroid of the training data. Points represents the predictions for 4000 randomly generated (uniform distribution) input data entries inside the limits of the training dataset. Points inside data region convex hull are presented in red. This is shown for the variables  Ped. Width ($\Delta$), $T_{e}(\psi_0)$ and $n_{e}(\psi_0)$.}
    \label{fig:error_random}
\end{figure*}

However, another important point of investigating the BNN architecture revolves around its potential to differentiate interpolation regions from extrapolation regions via its uncertainty predictions~\cite{aBNN_NCP-Hafner}. This study examines this capability using the epistemic uncertainty prediction, where a generally lower uncertainty is expected in the interpolation regions. Figure~\ref{fig:error_random} shows the relative predicted epistemic and aleatoric uncertainty for $\Delta$, $T_{e,ped}$, and $n_{e,ped}$ using randomly generated input data, as a function of the Euclidean distance of the input data to the centroid of the training input dataset. This distance is calculated after re-scaling these random samples with the same relation used to re-scale the input data distributions to have a mean of 0 and a standard deviation of 1. This is done to avoid introducing any bias on the distance metric due to the orders of magnitude differences spanned by the absolute values of the various input variables. It is also noted that some of the random input data points used may not resemble realistic operational plasma conditions, though this is not a requirement for demonstrating the characteristics of the uncertainty predictions under extrapolation. The training data centroid was chosen as the reference point as it should best represent the region which has the highest data density. Thus, as the distance increases, the epistemic error should increase as well because the neural network would have less information to train its predictive capability. 4000 input entries were randomly generated using two uniform distributions: the first one over the central 60 \% of the dataset and the second over the central 98 \%. This has been done intentionally to highlight both the points inside the training convex hull and those far from it.

Indeed, Figure~\ref{fig:rel_epi} demonstrates that when the distance increases, the epistemic uncertainty also increases notably. The random points that fall inside the convex hull of the training data region are distinguished in these plots. In the epistemic case, the uncertainty is expected to be low inside the hull and growing as the points move further out of it. In the aleatoric case, the uncertainty should be higher inside the hull and decrease as the point move further out of it. Indeed, this is roughly observed in Figure~\ref{fig:error_random}. As the neural network catches the variance of the training input data through the aleatoric uncertainty, going far from training data means low aleatoric uncertainty as determined by the target $s_j$ values chosen in this study. On the other hand, epistemic uncertainty reflects the precision of the neural network regarding the stochastic variance of the weights and biases with respect the training data, then out of the data region this precision should be much lower, due to the lack of information. Although the general behavior of the uncertainty is well captured in Figure~\ref{fig:error_random}, a 1D toy example can is provided in Section~\ref{subsec:phys_check} where it is more intuitively depicted while simultaneously verifying the model via a physical scan.

Additionally, in Appendix~\ref{sec:samp_unc} a study over the input uncertainties effect on the output result of EuroPED-NN has been done. It can be observed that in most cases the aleatoric uncertainty accounts for the input uncertainty.

\subsection{Independent physical analysis}
\label{subsec:phys_check}

In order to verify the model based on existing physical intuition about pedestal behaviours, this section compares the EuroPED-NN predictions against known empirical relations.

\subsubsection{$I_p$ scan}
\label{subsubsec:ip_scan}

Although the precise behaviour is dependent on many variables, it is generally expected that $n_{e,ped}$ grows when increasing $I_p$~\cite{AEHubbard_2000}. As $n_{e}(\psi_0)$ is one of the outputs of the model and $I_p$ is used in the calculation of the model input, $\mu$, given in Equation~\eqref{eq:mu}, this trend should also be present in EuroPED-NN. To verify this, a 1D scan is performed over $I_p$ while keeping the other inputs at a fixed value.

\begin{figure}[!ht]
    \centering
    \includegraphics[width=\linewidth]{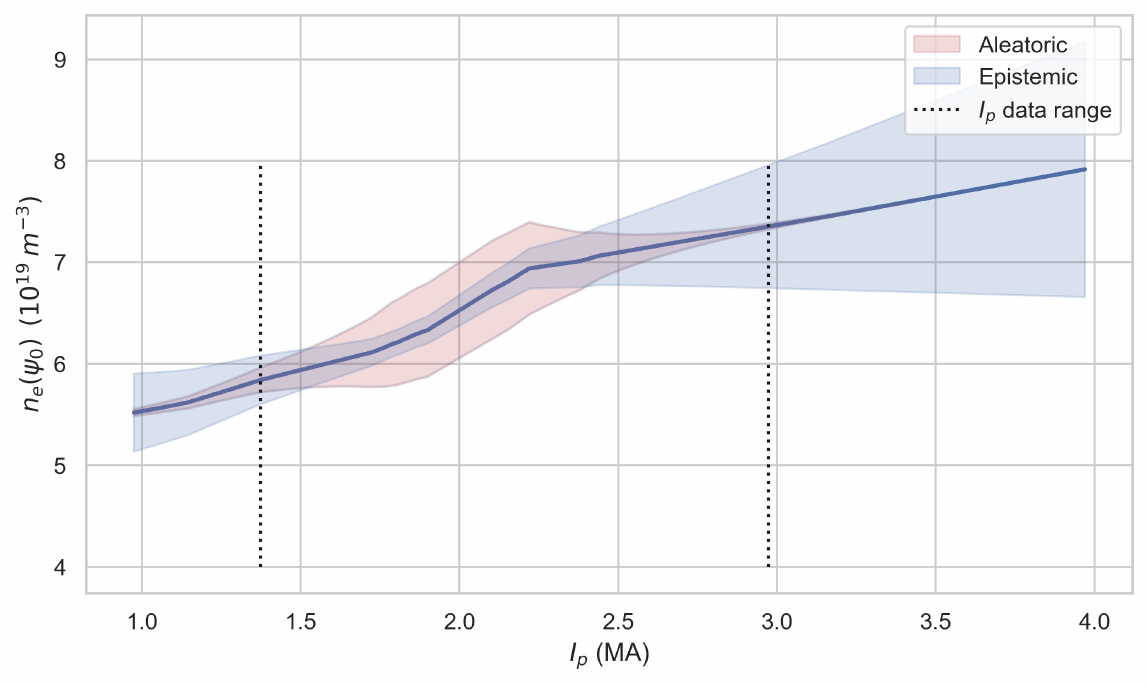}
    \caption{One dimesional scan for input variable $I_p$. Model prediction for $n_{e}(\psi_0)$ is shown for different values of $I_p$ and fixed values for the rest of the inputs. Bands for \emph{epistemic} and \emph{aleatoric} uncertainties are appreciated. The data region in $I_p$ is represented as the space between two vertical dashed lines.}
    \label{fig:1d_scan}
\end{figure}

Figure~\ref{fig:1d_scan} clearly shows that $n_{e}(\psi_0)$ increases with $I_p$, especially in the training data region between the dotted lines. While it is expected that not all input combinations in this scan correspond to physically achievable plasma scenarios, this trend was found to be present across a majority of random combinations of the other input parameters, of which the 1D scan plots are not shown for brevity. It can then be inferred that it extends to the subspace of physically-relevant parameters as well. Additionally, as mentioned in Section~\ref{subsec:Performance}, it is noticeable how the region populated by the training database exhibits significantly lower epistemic uncertainty and higher aleatoric uncertainty than the regions outside of it.

\subsubsection{$\Delta-\beta_{p,ped}$ relation}
\label{subsubsec:delta_beta}

As mentioned in Section~\ref{subsec:ModelInputOutput}, the semi-empirical relation between the pedestal width, $\Delta$, and the normalised plasma pressure, $\beta_{p,ped}$, is an important component of both EPED and EuroPED models. Thus, it is relevant to check whether the EuroPED-NN predictions replicate this relation. However, as the model outputs were reframed to provide the kinetic parameters at $\psi=\psi_0$ while retaining $\beta_{p}\!\left(\psi_0\right) \approx \beta_{p,ped}$ for purposes of consistency in integrated modelling, it is also important to illustrate that the surrogate model retains this feature by plotting $\Delta$ against $\beta_{p}\!\left(\psi_0\right)$. This is due to the fact that the outputs of EuroPED-NN are: the electron temperature prediction $T_{e}\!\left(\psi_0\right)$, electron density prediction $n_{e}\!\left(\psi_0\right)$ and the pedestal width $\Delta$, and all of them are included somehow in the plot. Due to the fact that $T_{e}\!\left(\psi_0\right)$ and $n_{e}\!\left(\psi_0\right)$ are needed for calculating $\beta_{p}\!\left(\psi_0\right)$. Moreover, to perform this calculation, given in Equation~\eqref{eq:beta}, some of the model inputs are required: plasma current, effective charge, triangularity, elongation and minor radius. These values are used to calculate $\beta_p$, which ultimately is a measure of the confinement, expressed as:
\begin{equation}
    \label{eq:beta}
    \beta_p=\frac{\left\langle p \right\rangle}{\left\langle B_{p} \right\rangle^2 / 2\mu_0}
\end{equation}
where $B_p$ is the poloidal component of the magnetic field, and the angle brackets represent flux-surface averages made around the poloidal angle.

Then this $\Delta-\beta_{p}\!\left(\psi_0\right)$ representation shows roughly the behavior of the model, as it uses all the output information and some input information. The similarities between the plot generated from predictions of EuroPED and from the reduced model predictions will help to verify the model with a physics-based representation and to measure the performance of the surrogate model.


\begin{figure*}[tb]
    \centering
    \begin{subcaptionblock}{6cm}
        \centering
        \includegraphics[width=\linewidth]{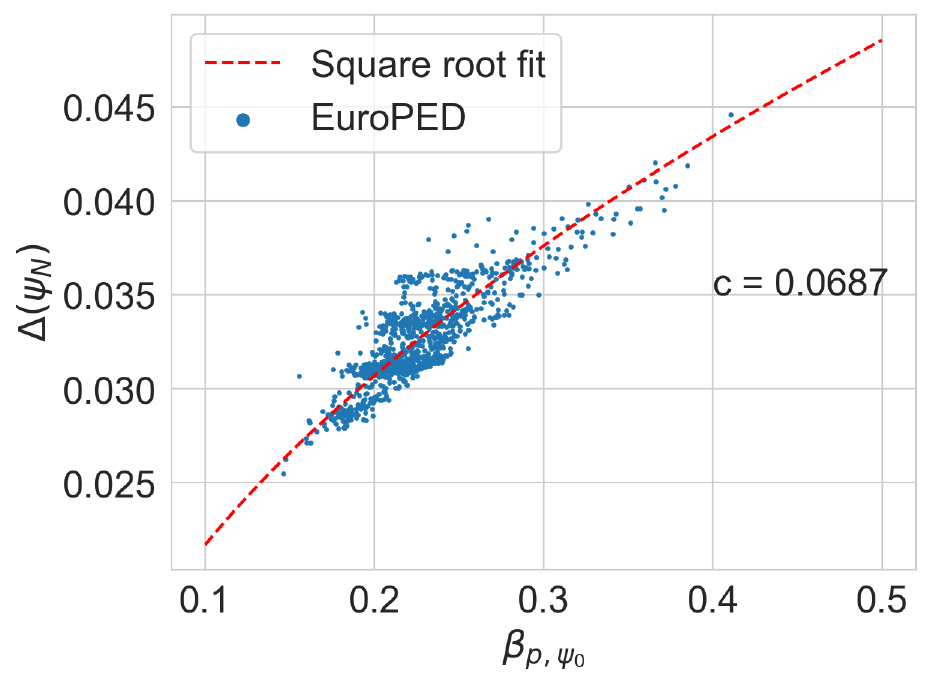}
        \caption{EuroPED dataset\\
        }
        \label{fig:beta_delta_europed_db}
    \end{subcaptionblock}%
    \begin{subcaptionblock}{6cm}
        \centering
        \includegraphics[width=\linewidth]{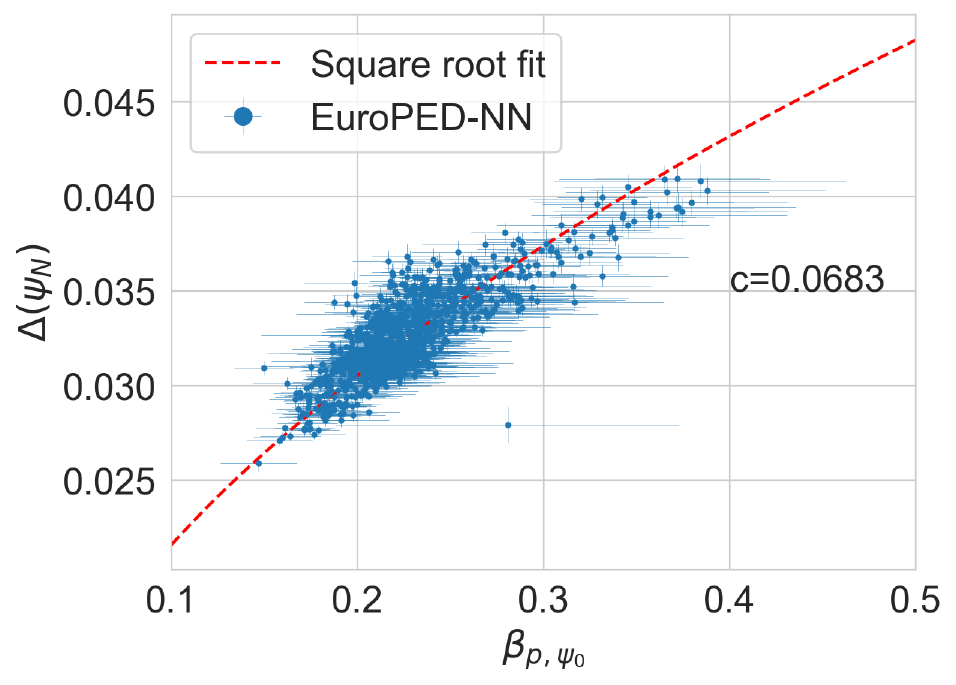}
        \caption{EuroPED-NN predictions on dataset points}
        \label{fig:beta_delta_sur_epi}
    \end{subcaptionblock}%
    \begin{subcaptionblock}{6cm}
        \centering
        \includegraphics[width=\linewidth]{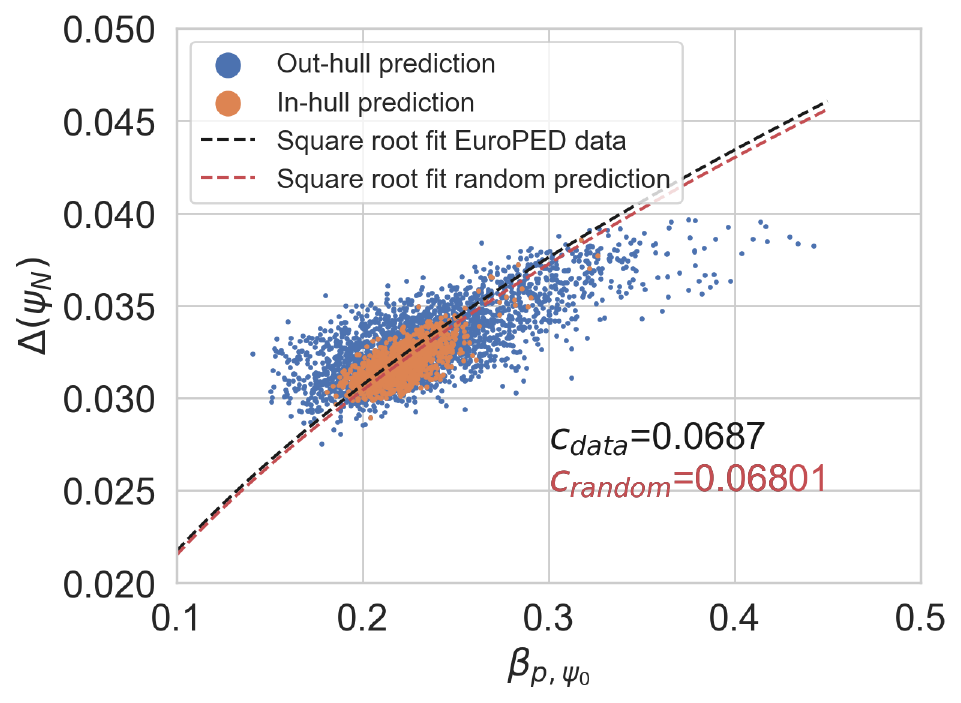}
        \caption{EuroPED-NN predictions on random points}
        \label{fig:beta_delta_rand}
    \end{subcaptionblock}
    \caption{Data from various sources represented in pedestal-width ($\Delta$)-$\beta_{p}\!\left(\psi_0\right)$ space, with errorbars showing the epistemic uncertainty. Predictions (blue dots) are fitted to show the semi empirical relation $\Delta=c\sqrt{\beta_{p,ped}}$ (square root fit in dashed red). The pedestal width is measured in normalised magnetic flux units $\psi_{\text{pol}}$ and $\beta_{p}\!\left(\psi_0\right)$ is a dimensionless quantity. In \textbf{(c)}, prediction points inside the training convex hull are shown in orange.}
    \label{fig:beta_delta_europed}
\end{figure*}

Figure~\ref{fig:beta_delta_europed_db} shows the behaviour of the original EuroPED original predictions for JET-ILW experiment database in the $\Delta-\beta_{p}\!\left(\psi_0\right)$ space. The original square root relation used in EuroPED and EPED models can be appreciated as: $\Delta=c\sqrt{\beta_{p,ped}}$. This relation was found in the tokamak DIII-D to be held with $c=0.076$~\cite{EPED_paper}. However, as EuroPED-NN gives the outputs at $\psi=\psi_0$ and not at top pedestal for convenience purposes, the relation cannot be hold in the exact same way, then some scatter around the square root fit can be found for both EuroPED and EuroPED-NN predictions. This is a result from the discrepancy between $\beta_{p,ped}$ and $\beta_{p}\!\left(\psi_0\right)$.

Figure~\ref{fig:beta_delta_sur_epi} show the EuroPED-NN predictions with epistemic uncertainty, across both the training and test datasets from the pedestal database. This means that the plotted uncertainties should correspond to the interpolation region, where the surrogate model is expected to provide decent predictions.

In Figure~\ref{fig:beta_delta_rand}, the EuroPED-NN predictions for random inputs inside the range of the dataset inputs are shown. 4000 input entries have been generated using two uniform distributions, one using the central 90\% range of the dataset and the other the central 60\%, this has been done to adequately populate the $\Delta-\beta_{p}\!\left(\psi_0\right)$ space. A convex hull using the training data has been created in order to check which random points fall inside of it. It can be appreciated in the plot how the points inside the convex hull are following the square root fit with higher precision than the rest. Then the $\Delta-\beta_{p}\!\left(\psi_0\right)$ relation has been caught and generalized adequately to the training space by EuroPED-NN.

Overall, it is clear that the EuroPED-NN predictions in the $\Delta-\beta_{p}\!\left(\psi_0\right)$ space is quite similar to the original one in Figure~\ref{fig:beta_delta_europed_db}, moreover the random points seem to follow the trend, specially those inside the training convex hull, so the model is capturing the behavior reasonably well.

\subsection{Contextualization in data-driven pedestal models}\label{sec:context}

Data-driven methods, in particular neural networks, have been applied to predict plasma pedestal quantities in the past. To put EuroPED-NN in its context a comparison will be carried out with EPED1-NN \cite{aTGLFEPEDNN-Meneghini} and PENN \cite{Gillgren_2022} models, to know which are their main attributes and where EuroPED-NN stands in relation to them.

Regarding EPED1-NN, the model was developed to replicate EPED pedestal model \cite{EPED_paper}. The inputs of the model are the inputs of EPED: $n_{e,ped}$, $Z_{eff}$, $\beta_N$, $I_p$, $B_t$, $a$, $R$, $\kappa$ and $\delta$. The outputs are pedestal pressure $p_{ped}$ and pedestal width. EPED1-NN is built using a regular feed-forward neural network, and its accuracy achieves values around $R^2=0.987$ and $R^2=0.964$, for the respective outputs, when compared with EPED predictions as described in \cite{aTGLFEPEDNN-Meneghini}.

On the other hand PENN model is intended to replicate experimental data from JET-ILW pedestal database~\cite{aJETExperimentalPedestalDatabase-Frassinetti}. The inputs of the model are $Z_{eff}$, $\beta_N$, $I_p$, $B_t$, $a$, $\kappa$, $\delta_{up}$, $\delta_{low}$, $\kappa$, NBI Power, Total Power, plasma volume and $q_{95}$. Where $\delta_{up}$, $\delta_{low}$ and $q_{95}$ stands for upper triangularity, lower triangularity and q value at $\psi =0.95$ respectively. PENN is also build using a regular feed-forward neural network and the output variables are the pedestal density $n_{e,ped}$ and temperature $T_{e,ped}$, for which the accuracy in test data when compared with experimental data is $R^2=0.91$ and $R^2=0.93$, respectively.

Therefore, we can compare both models characteristic with EuroPED-NN, where the set of inputs (explained in the Section~\ref{subsec:ModelInputOutput}) differ from both EPED1-NN and PENN, then the usage cases are different, as happens between EPED and EuroPED. Depending on the available data one model can be more suitable than the other. The accuracy is reasonably good in all of them, including EuroPED-NN with $R^2_{\Delta,test}=0.846$, $R^2_{T_e,test}=0.874$, and $R^2_{n_e,test}=0.970$ (see Figure \ref{fig:perf}). However, this accuracy is quoted against the respective datasets of each model, meaning that the insights that can be extracted from this comparison are limited. Regarding uncertainty analysis, only EuroPED-NN is an uncertainty-aware model inherently providing its corresponding model and data uncertainty estimates. Regarding the model applicability, it is known that neither EPED nor EuroPED includes the effect of plasma resistivity. Thus, the empirical model PENN could be a better option when predicting the properties of a more resistive plasma. However, to explore the physics and behavior of the pedestal models in a fast way, both EPED1-NN and EuroPED-NN are also applicable.

\subsection{Extrapolation of EuroPED-NN to AUG data}
\begin{figure*}
    \centering
    \includegraphics[width=\linewidth]{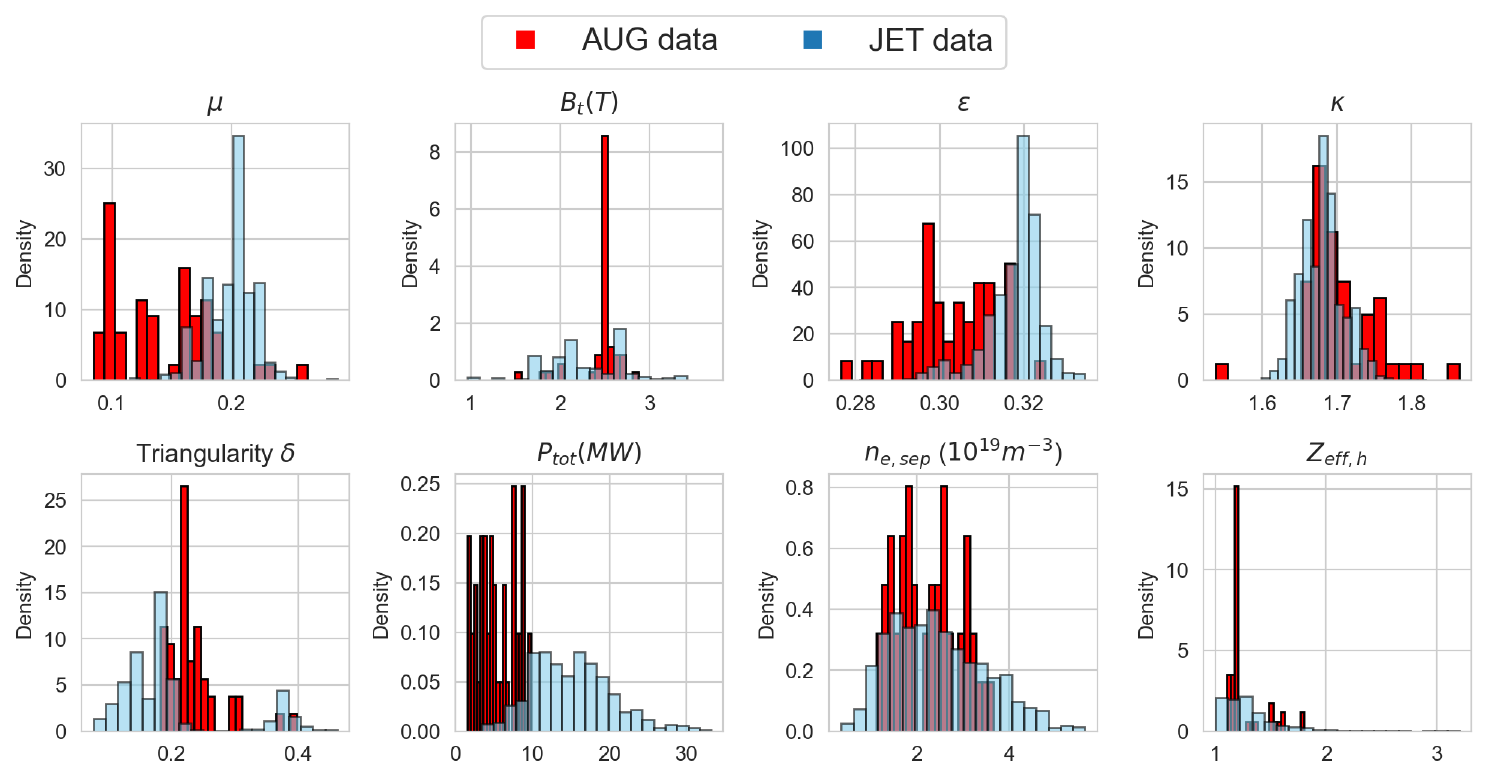}
    \caption{Input data histograms for EuroPED-NN from the JET dataset used to train and test in blue, and from AUG data in red, obtained from~\cite{aJETExperimentalPedestalDatabase-Frassinetti} and from \cite{Luda_2020,Luda2021} respectively.}
    \label{fig:hist_JET_AUG}
\end{figure*}

\begin{figure*}[tb]
    \centering
    \begin{subcaptionblock}{5.8cm}
        \centering
        \includegraphics[width=\linewidth]{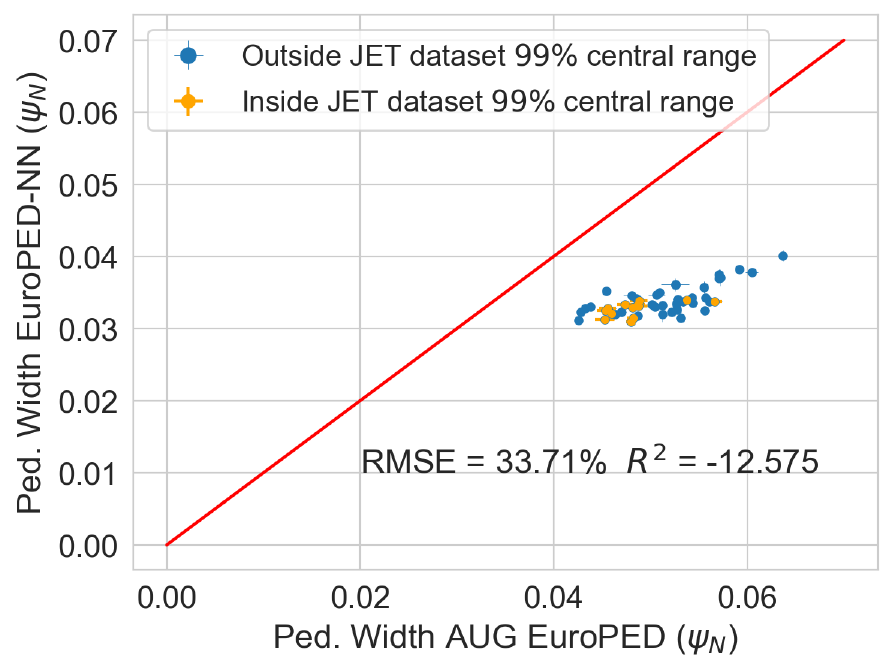}
        \caption{$\Delta$}
        \label{fig:delta_europed_AUG}
    \end{subcaptionblock}%
    \begin{subcaptionblock}{5.8cm}
        \centering
        \includegraphics[width=\linewidth]{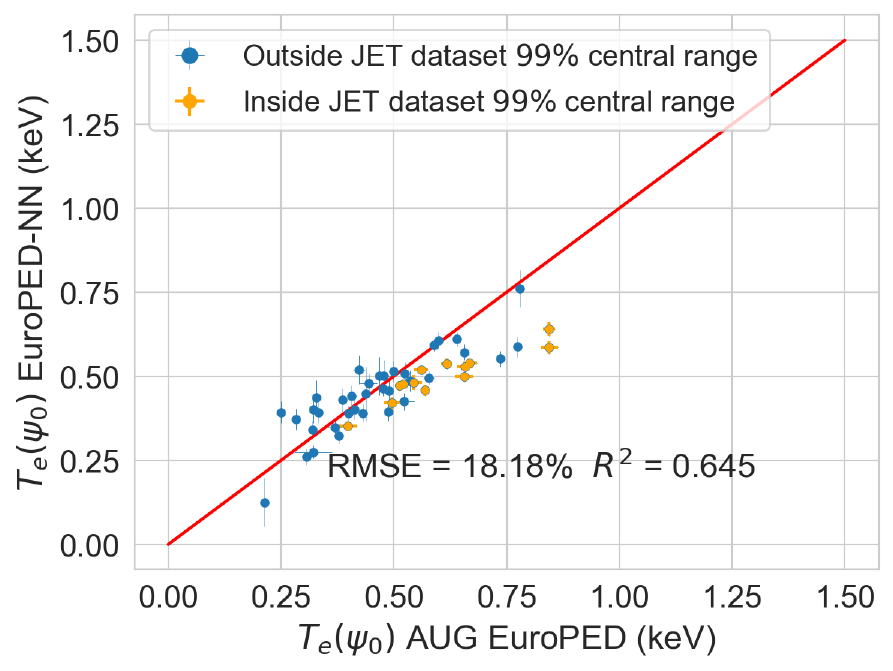}
        \caption{$T_{e}(\psi_0)$}
        \label{fig:teped_europed_AUG}
    \end{subcaptionblock}
    \begin{subcaptionblock}{5.8cm}
        \centering
        \includegraphics[width=\linewidth]{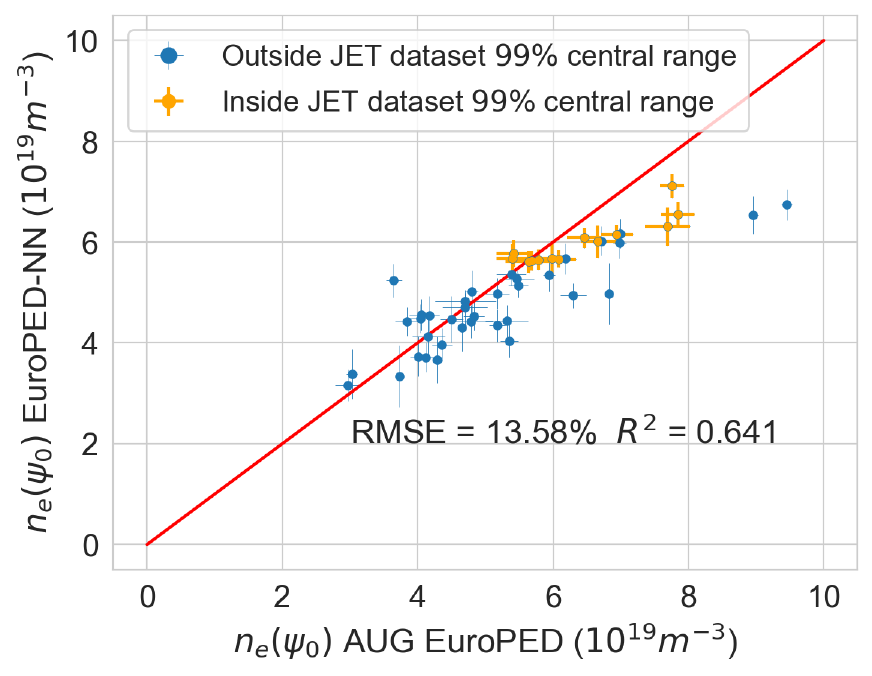}
        \caption{$n_{e}(\psi_0)$}
        \label{fig:neped_europed_AUG}
    \end{subcaptionblock}%
    \caption{EuroPED-NN performance in AUG data for its three output variables. Predictions corresponding to input points inside and outside 99 \% central JET data range are shown in orange and blue, respectively. Epistemic and aleatoric uncertainties are displayed in vertical and horizontal axis respectively.}
    \label{fig:perf_europed_AUG}
\end{figure*}

Finally, an experimental pedestal database from ASDEX-Upgrade (AUG) tokamak~\cite{Luda2021,Luda_2020} with 50 shots, and its subsequent EuroPED predictions were available. This data is used to compare the EuroPED-NN predictions with the EuroPED evaluations. As the AUG database parameters lie outside the hull of JET parameters, the exercise constitutes an extrapolation of the model. This will test the adaptability of EuroPED-NN out of JET database values. The distribution of input values from the 50 AUG shots can be compared in Figure \ref{fig:hist_JET_AUG} with the JET database. The biggest difference between both of them are found in $\mu$, $\epsilon$, $P_{tot}$ and triangularity $\delta$ variables.\\
In the Figure \ref{fig:perf_europed_AUG} the EuroPED-NN predictions performance in AUG data is displayed for the three output variables. The points are split between those whose input points lie inside the central 99\% of JET database range, and those who doesn't. This central 99\% boundary was chosen to effectively exclude outliers from artificially expanding the range limits. It can be appreciated how most of the points are outside this range, so it is therefore an extrapolating exercise. For $T_e(\psi_0)$ and $n_e(\psi_0)$ the match with the EuroPED predictions is reasonable with $R^2=0.645$ and $R^2=0.641$, respectively. However, in the case of the pedestal width, the extrapolation does not look to have succeeded. The reason for this is that the AUG predictions were run using a KBM constant of 0.09 (instead of the standard 0.076) as AUG has found that the standard prediction gives too narrow values. For this reason, the pedestal width variable cannot be extrapolated, while the rest variables perform much better.\\
The fact that EuroPED-NN performance in AUG data is reasonable good in the pedestal height quantities, even though the model is extrapolating, might mean that EuroPED-NN is adequately catching the trend between the input data and the pedestal quantities, rather than just adapting to the training data. However, to increase its accuracy and generalization power, future work training EuroPED-NN with data from other machines is encouraged.
\section{Extension of the methodology to experimental data and extrapolation to AUG experimental data.}
\label{sec:exp_model}

\begin{figure*}[!h]
    \centering
    \begin{subcaptionblock}{6cm}
        \centering
        \includegraphics[width=\linewidth]{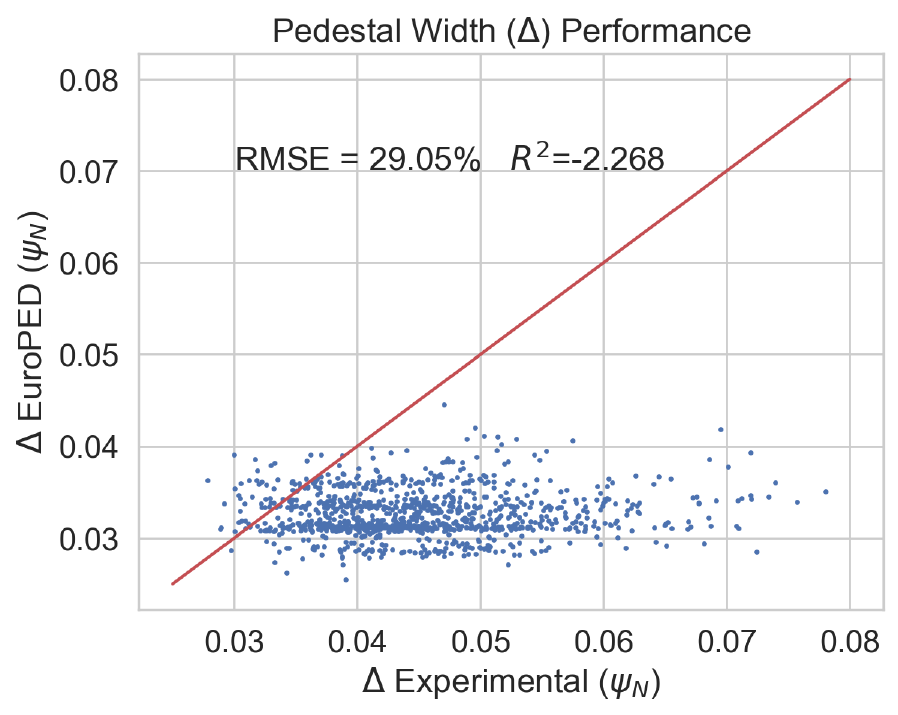}
        \caption{$\Delta$}
        \label{fig:delta_perf_europed}
    \end{subcaptionblock}%
    \begin{subcaptionblock}{6cm}
        \centering
        \includegraphics[width=\linewidth]{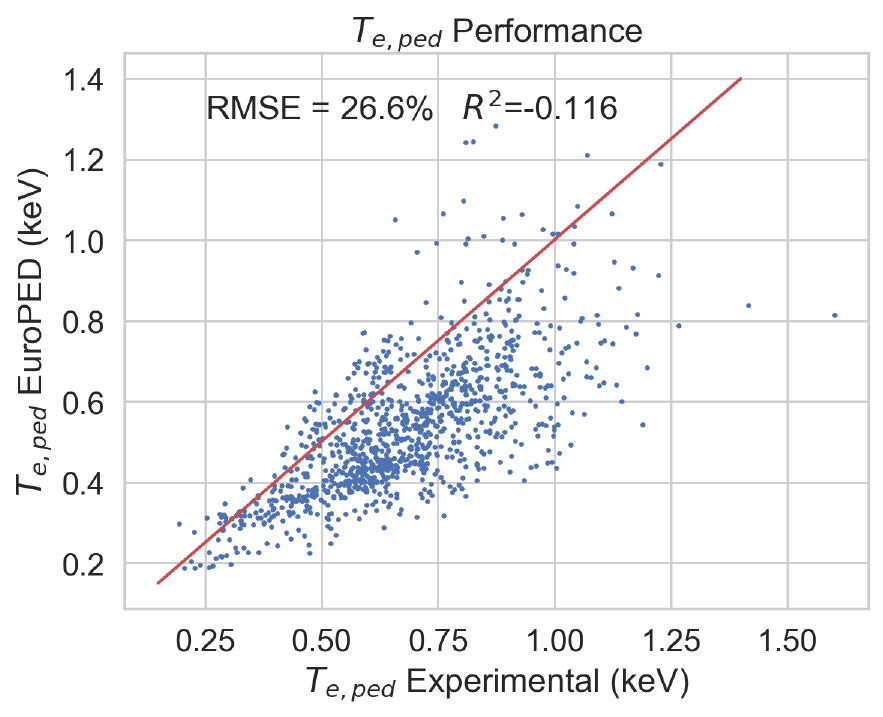}
        \caption{$T_{e,ped}$}
        \label{fig:teped_perf_europed}
    \end{subcaptionblock}%
    \begin{subcaptionblock}{6cm}
        \centering
        \includegraphics[width=\linewidth]{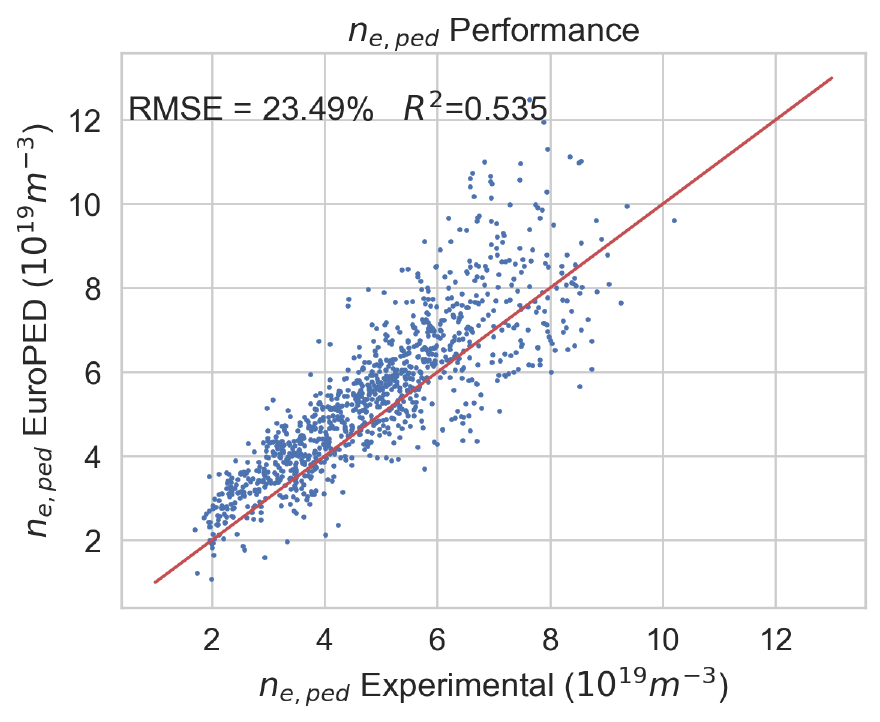}
        \caption{$n_{e,ped}$}
        \label{fig:neped_perf_europed}
    \end{subcaptionblock}
    \caption{EuroPED predictions performance compared with experimental values for its three output parameters.}
    \label{fig:perf_europed_exp}
\end{figure*}

\begin{figure*}[!h]
    \centering
    \begin{subcaptionblock}{6cm}
        \centering
        \includegraphics[width=\linewidth]{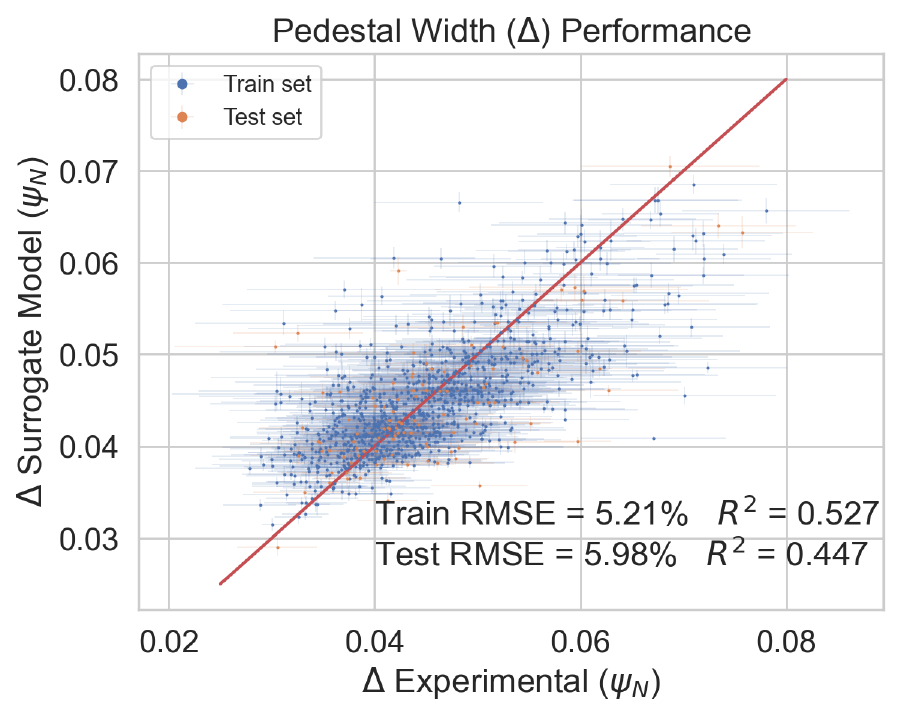}
        \caption{$\Delta$}
        \label{fig:delta_perf_exp}
    \end{subcaptionblock}%
    \begin{subcaptionblock}{6cm}
        \centering
        \includegraphics[width=\linewidth]{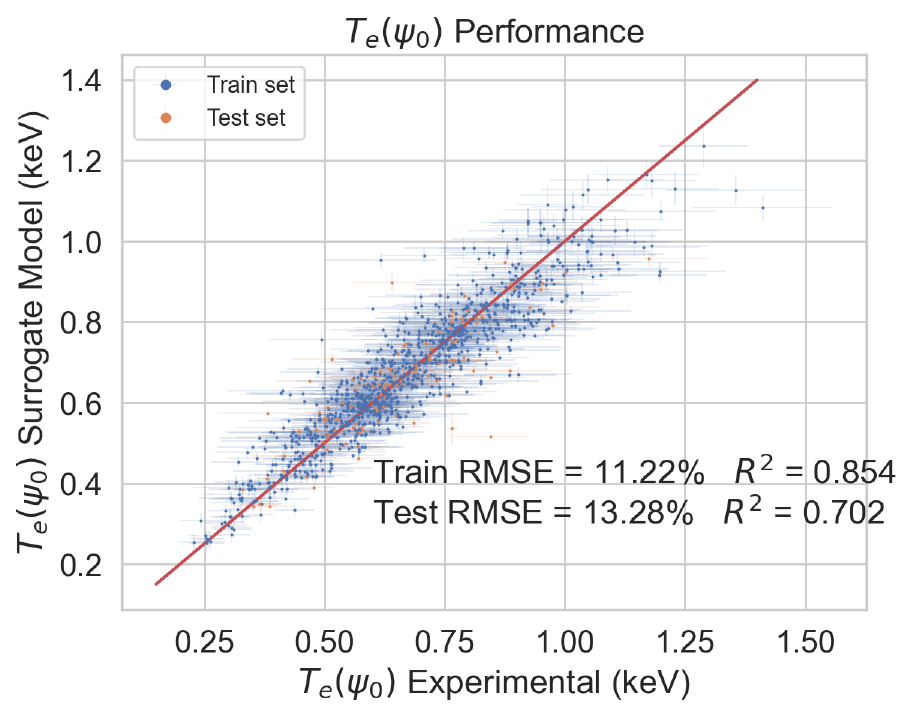}
        \caption{$T_{e,ped}$}
        \label{fig:teped_perf_exp}
    \end{subcaptionblock}%
    \begin{subcaptionblock}{6cm}
        \centering
        \includegraphics[width=\linewidth]{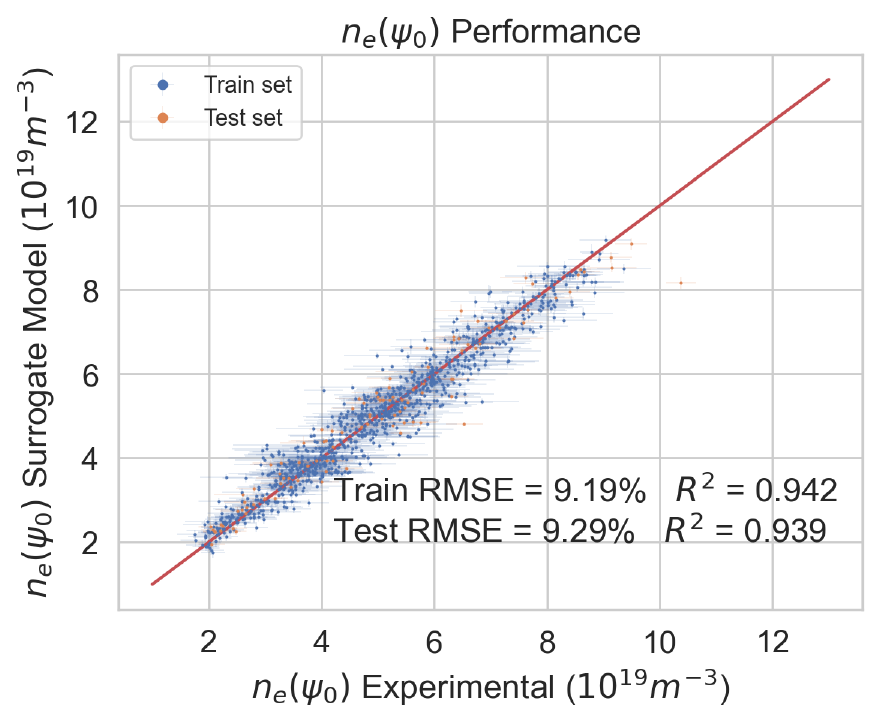}
        \caption{$n_{e,ped}$}
        \label{fig:neped_perf_exp}
    \end{subcaptionblock}
    \caption{Experimental surrogate model predictions for its three output parameters compared with the respective experimental data. Epistemic and aleatoric uncertainties are displayed in vertical and horizontal axis respectively.}
    \label{fig:perf_exp_nn}
\end{figure*}

During the training of the EuroPED-NN model, it was found that some of the EuroPED output predictions do not appear to align with the experimental data. Figure~\ref{fig:perf_europed_exp} shows a comparison between the experimental data and EuroPED-predicted outputs in the database used in this study, where it becomes evident that the prediction points do not follow the 1:1 line. While the objective of this study was to train a surrogate model for EuroPED, its accuracy is ultimately constrained to the accuracy of EuroPED itself. To push the topic of general pedestal surrogate modelling further, an experimental model was constructed using the same EuroPED inputs while swapping the output data to their corresponding experimental values. By employing the same BNN-NCP structure to create this upgraded surrogate model, it exhibits a much better alignment for the same variable, as illustrated in Figure~\ref{fig:perf_exp_nn} if we compare it with Figure~\ref{fig:perf_europed_exp}. Then this experimental model clearly outperforms EuroPED in the JET-ILW database.


However, when comparing Figures~\ref{fig:perf} and \ref{fig:perf_exp_nn}, it becomes evident that the accuracy of this ``experimentally-enhanced" surrogate model is significantly lower than that of the EuroPED surrogate model. This suggests that the chosen set of input variables for EuroPED may be insufficient to sufficiently characterize the pedestal physics contained in the dataset, based on its inability to obtain a similarly high model performance, as determined by its RMSE and $R^2$. Guided by recent pedestal physics literature, an attempt was made to include the normalized plasma collisionality, $\nu^*$, as an input variable. However, these efforts have proven inadequate in explaining why the model performance cannot be recovered between the models. Nevertheless, this marks a promising initial study into the development of a sufficient set of parameters to fully describe pedestal physics and further research in this direction is strongly recommended.

It may seem reasonable to compare this experimental pedestal model with the previously described PENN model~\cite{Gillgren_2022} as both of them are based on experimental data, although the set of input variables differ from each other. As mentioned, PENN model has an accuracy in test data of $R^2=0.91$ and $R^2=0.93$, for $n_{e,ped}$ and $T_{e,ped}$ respectively, while the experimental model developed in this section is associated with test set accuracy of $R^2=0.94$ and $R^2=0.70$. Then it can be seen how the experimental model performs better for $n_{e,ped}$, while PENN performs significantly better for $T_{e,ped}$. The likely explanation lies in the input parameters of each model. The model utilized in this study benefits from $n_{e,sep}$ as input parameter, which is closely related with $n_{e,ped}$. In contrast, the PENN model boasts a greater number of inputs, enhancing its ability to provide more comprehensive information for accurate predictions. It is worth noting that the experimental model inputs have been purposefully selected to correspond with the EuroPED inputs, so an input parameter analysis has not been carried out.

\begin{figure*}[!ht]
    \centering 
    \begin{subcaptionblock}{6cm}
        \centering
        \includegraphics[width=0.9\linewidth]{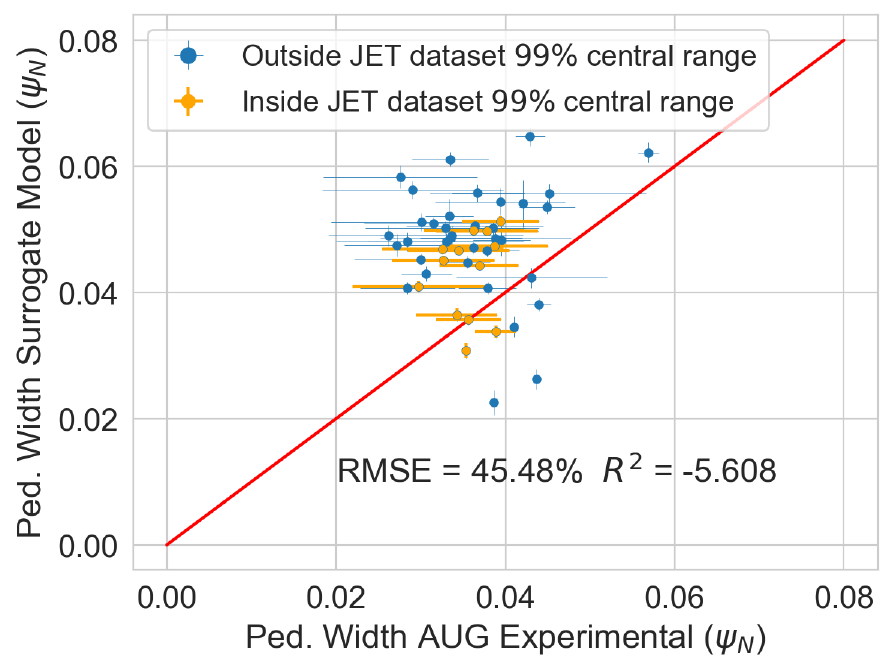}
        \caption{$\Delta$}
        \label{fig:teped_AUG_exp}
    \end{subcaptionblock}%
    \begin{subcaptionblock}{6cm}
        \centering
        \includegraphics[width=0.9\linewidth]{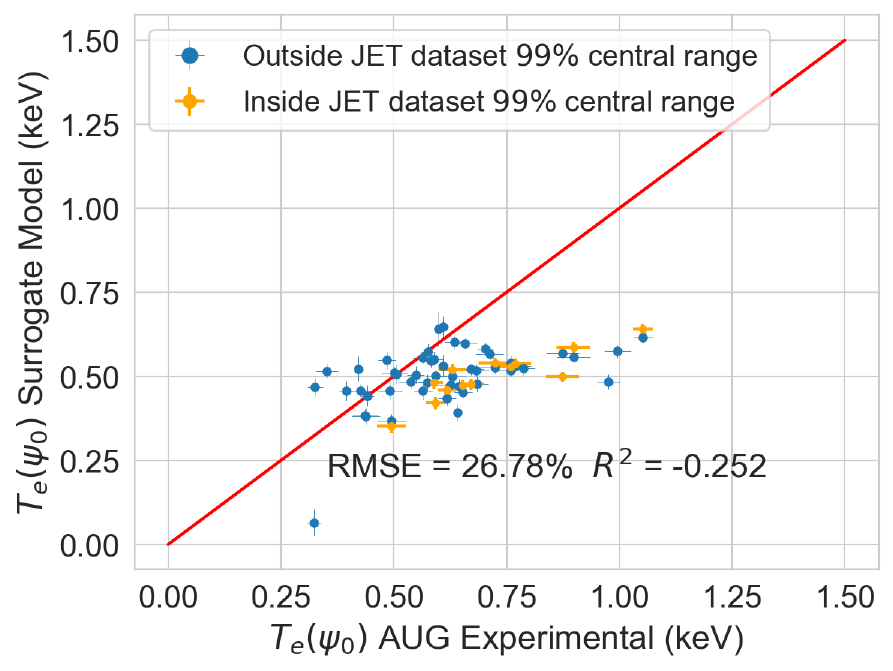}
        \caption{$T_{e,ped}$}
        \label{fig:teped_AUG_exp}
    \end{subcaptionblock}%
    \begin{subcaptionblock}{6cm}
        \centering
        \includegraphics[width=0.9\linewidth]{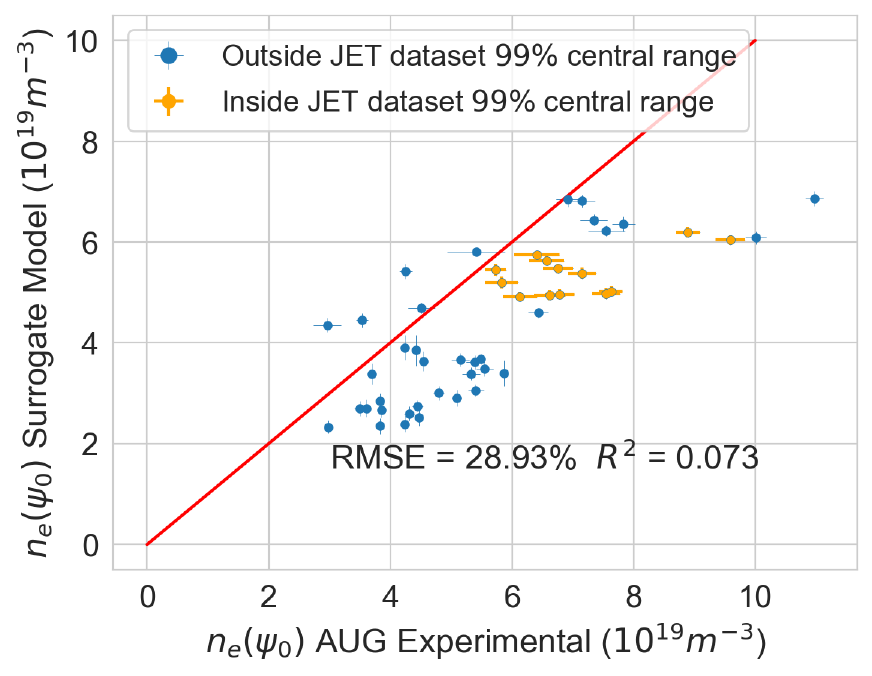}
        \caption{$n_{e,ped}$}
        \label{fig:neped_AUG_exp}
    \end{subcaptionblock}
    \caption{Experimental surrogate model predictions for pedestal width, $T_{e}(\psi_0)$ and $n_{e}(\psi_0)$ compared with the respective experimental data from AUG tokamak. Points are split regarding the position of the input values with respect to the JET training database. Epistemic and aleatoric uncertainties are displayed in vertical and horizontal axis respectively}
    \label{fig:AUG_exp}
\end{figure*}

Finally, the experimental pedestal database from ASDEX-Upgrade (AUG) tokamak~\cite{Luda2021,Luda_2020} will be used to compare the experimental model predictions with the data from AUG. Again, as the AUG database parameters lie outside the hull of JET parameters, the exercise constitutes an extrapolation of the model. Also note that EuroPED model does not play any role in this exercise, as the experimental model has been trained exclusively with experimental data, not with the subsequent EuroPED evaluations, although both have the same set of inputs. In Figure~\ref{fig:AUG_exp}, the accuracy of the experimental model in the AUG cases is represented for pedestal width, $T_{e,ped}$ and $n_{e,ped}$, the prediction points are split between those whose input values fall inside the central 99\% of JET training database range and those that do not. It can be inferred from the accuracy metrics that the experimental NN model performs significantly worse in experimental AUG pedestal values, although in the plots it can be seen how in the case of $T_{e,ped}$ and  $n_{e,ped}$ it roughly gets the trend of the 1:1 line. Interestingly, the model does not return pedestal values corresponding to the AUG data even for the AUG points which lie within the JET training parameter space. A possible explanation for this can be due to the fact that the training dataset contains $\sim$1000 entries, which may not be enough for the NN to learn the necessary relations to generalize the problem in the experimental case, even within its own training space. Another explanation could be the election of input parameters, which, as commented, may not be enough to recover the experimental pedestal values. A more thorough investigation into the input variables selection or the incorporation of other physical constraints is recommended to improve the results of cross-machine comparisons, especially if the databases used cannot be significantly expanded for future studies.

\section{Conclusion}
\label{sec:Conclusion}


In this study, it has been shown that EuroPED-NN, trained using the BNN-NCP technique, is able to reproduce the results and behavior of EuroPED JET database with good accuracy, whose mean prediction provides similar performance to that of a simple feed-forward neural network, showing performances. Obtaining errors under 7 \% for $T_{e}\!\left(\psi_0\right)$ and $n_{e}\!\left(\psi_0\right)$ and under 3 \% for pedestal width. However the use of this BNN-NCP technique has allowed to identify extrapolation and interpolation regions via the epistemic and/or aleatoric uncertainties, as the epistemic uncertainty generally grows with distance from dataset convex hull, while the aleatoric uncertainty falls. On top of that, the use of these aleatoric and epistemic uncertainties allows the model to be uncertainty aware in the predictions, so that the prediction's confidence can be known.

EuroPED-NN has also been tested using physical reasoning. First, a one-dimensional scan calculating $n_{e}\!\left(\psi_0\right)$ for different values of $I_p$ has been performed proving the expected growth of $I_p$, while showing the behavior of the prediction's uncertainty. Second, the $\Delta-\beta_{p,ped}$ semi-empirical relation included inside EuroPED has been tested to hold in EuroPED-NN, using both JET dataset inputs and random inputs, so that the relation is correctly generalized.

Finally, after explaining the method, building the surrogate model and performing the posterior analysis, it can be said that the EuroPED-NN is completely functional in JET-ILW data. Additionally, the extrapolation of EuroPED-NN to AUG tokamak data has been tested, obtaining reasonable accuracy for $T_e(\psi_0)$ and $n_e(\psi_0)$ of around $R^2=0.64$. However, the pedestal width accuracy was much lower as EuroPED was run in AUG data using a different constant for the KBM constraint. In general, better performance in this exercise would typically require a bigger set of training data and/or training data from other tokamaks. Future work is encouraged in this direction.

To sum up, EuroPED-NN predictions closely match EuroPED predictions in JET input data, with errors of less than 7\%  for both electron temperature ($R^2=0.874$) and electronic density ($R^2=0.970$), and less than 3\% for pedestal width ($R^2=0.846$). The model identifies the areas where the prediction uncertainty is higher and lower, which is helpful for understanding its reliability. Furthermore, the model's performance was validated by comparing its predictions with known physical principles, demonstrating its understanding of some plasma physics principles involved in EuroPED.

The surrogate model EuroPED-NN is mainly intended to be used in plasma control, scenario development and physics exploration of pedestal plasma. The advantage with respect to EuroPED model is not only the fast computation but also the uncertainty awareness. While this uncertainty is split in model uncertainty (epistemic) and data uncertainty (aleatoric) for further clarity. EuroPED-NN proves to be a valuable tool for various studies that can leverage any of this features. The model architecture and weights are located in this GitHub link~\url{https://github.com/alexpanera/EuroPED-NN}, any feedback on the use of the model will be appreciated.

In Section~\ref{sec:exp_model}, the same method and architecture developed for EuroPED-NN was also applied to experimental data using the same JET database, proving the adaptability of the method and bridging the gap between EuroPED predictions and experimental data, a challenge that EuroPED-NN inherently cannot address. While its performance surpassed that of EuroPED when compared with JET experimental data, it did not match the performance of EuroPED-NN compared to EuroPED. Furthermore, this experimental model was tested with AUG data, revealing its limited extrapolating power due to its low performance in this different tokamak. This could be due to the low number of training data points ($\sim 1000$) and/or the election of the EuroPED input variables. Future work is encouraged towards building a pedestal experimental model that can be generalised to various machines, likely using data from different machines.

In the Appendix~\ref{sec:samp_unc}, some additional analysis on EuroPED-NN uncertainties is performed, and in Appendix~\ref{sec:eped} the same methodology described has been implemented in the EPED model to prove again the adaptability of the method.

Furthermore, the method developed in this study (multidimensional BNN-NCP), based on a previous study detailing its methodology~\cite{aBNN_NCP-Hafner}, can be extended to build uncertainty aware models even beyond plasma physics, offering opportunities in diverse fields. In essence, the multidimensional BNN-NCP model with modified architecture that has been developed in this study holds immense potential for both scientific and non-scientific applications, by enabling the construction of precise uncertainty aware neural network models across multiple dimensions, improving the speed of regular Bayesian neural networks.

\section*{Acknowledgements}
This work has been carried out within the framework of the EUROfusion Consortium, funded by the European Union via the Euratom Research and Training Programme (Grant Agreement No 101052200 — EUROfusion). Views and opinions expressed are however those of the author(s) only and do not necessarily reflect those of the European Union or the European Commission. Neither the European Union nor the European Commission can be held responsible for them.
In particular, the work has been carried out within the framework of ENR-MOD.01.FZJ project.

\section*{Data availability}
Raw experimental data were generated at the Joint European Torus (JET) facility. Derived data and simulation results supporting the findings of this study are available from the corresponding author upon reasonable request.

\printbibliography

\appendix
\newpage
\section{OOD Hyperparameters}
\label{sec:tables}

This section provides some data for the OOD related hyperparameters referenced in the main body. These values were specifically chosen for this study based on the training attempts and expert reasoning of the authors. They must be reevaluated when applying the BNN-NCP technique to other applications.

In the case of the OOD sampling, as mentioned in the Section~\ref{subsec:HyperparameterTuning}, the width of OOD region for input sampling was chosen to be 2 times the standard deviation of the dataset input variables. This range resulted wide enough to get the desired results.

Regarding the target epistemic uncertainty $\mathbf{\Sigma_y}$, the value chosen in the three output variables was $\mathbf{\Sigma_y}=0.001$ in normalised units. The target aleatoric uncertainty was also chosen to be small, $\mathbf{s_y}=0.0001$ also in normalised units.

\section{Input uncertainty study in EuroPED-NN}
\label{sec:samp_unc}

In the methodology developed in this work two types of uncertainties are distinguished, epistemic and aleatoric. However, it is interesting to study how the input uncertainties are translated into the output uncertainties propagating them through the EuroPED-NN model. Those input uncertainties are estimated from the measurement signals and the values appear in Table~\ref{tab:sigma_x}.

\begin{table}[htb]
\centering
\begin{tabular}{l|c|}
\cline{2-2}
                                              & \multicolumn{1}{l|}{\textbf{$\sigma_{x}$}} \\ \hline
\multicolumn{1}{|l|}{$\mu$}                   & 0.02                                       \\ \hline
\multicolumn{1}{|l|}{$B_t$ (T)}               & 0.001                                      \\ \hline
\multicolumn{1}{|l|}{$\epsilon$}              & 0.01                                       \\ \hline
\multicolumn{1}{|l|}{$\kappa$}                & 0.01                                       \\ \hline
\multicolumn{1}{|l|}{Triangularity $\delta$}                & 0.01                                       \\ \hline
\multicolumn{1}{|l|}{$P_{tot}$ (MW)}          & 0.5                                        \\ \hline
\multicolumn{1}{|l|}{$n_{e,sep}$ ($10^{19}m^{-3}$)} & 0.7                                        \\ \hline
\multicolumn{1}{|l|}{$Z_{eff}$}               & 0.05                                       \\ \hline
\end{tabular}%
\caption{Values for the expected uncertainty in the input data.}
\label{tab:sigma_x}
\end{table}

The method to develop this study was similar to the procedure used in OOD sampling. The dataset input values were modified summing to each a sample of 1000 values generated by the distribution $\mathcal{N}\!\left(0,\sigma_x\right)$. Then this 1000 points generated with just one input entry are passed through EuroPED-NN, from this 1000 values for each of the outputs we obtain the standard deviation, which correspond with the input uncertainty translated to the output.

\begin{figure}[htb]
    \centering
    \includegraphics[width=\linewidth]{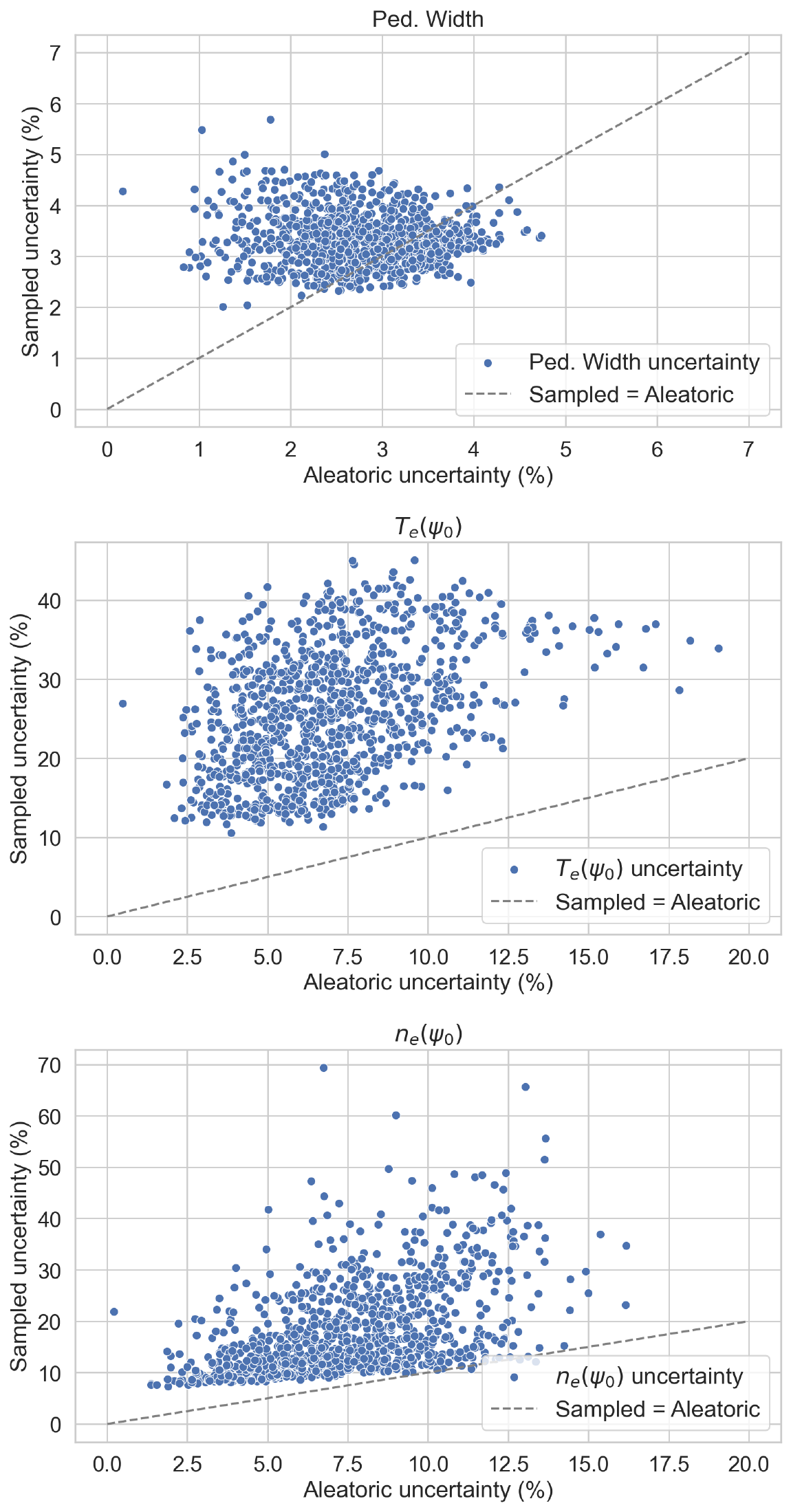}
    \caption{Comparison between sampled uncertainty coming from the input uncertainties and aleatoric uncertainty for all dataset points. }
    \label{fig:sampled_unc}
\end{figure}

Figure~\ref{fig:sampled_unc} shows the results of this study comparing the sampled uncertainty with the aleatoric uncertainty for all the points in the dataset. Aleatoric uncertainty was chosen for comparison due to the closest relation with the data variation, on top of that epistemic uncertainty was consistently lower than both aleatoric and sampled uncertainty. It is noticeable how in the output variables $n_{e}\!\left(\psi_0\right)$ and $T_{e}\!\left(\psi_0\right)$ the aleatoric uncertainty accounts for all the sampled uncertainty, serving as a lower bound. In the case of $\Delta$ both aleatoric and sampled uncertainty are in the same order. It is expected that the aleatoric uncertainty captures the input uncertainties translated to the outputs, as the aleatoric uncertainty catches the variance in the data, which is highly related with the measurements uncertainty. 

Note that the $n_{e,sep}$ uncertainty dominates the results, due to the large $\sigma_x$ values, as shown in Table~\ref{tab:sigma_x}, and due to the importance of the variable $n_{e,sep}$ in the calculation of the final output. To test impact of this choice, the sampling was repeated with lower values of $\sigma_x$ for $n_{e,sep}$, obtaining significantly lower sampled uncertainties.

Including the effect of the input uncertainty into the methodology of the neural network would be a desired feature for the model. In this work some attempts were performed, however the result was not satisfactory in most cases. Then it is suggested for future work.

\section{EPED surrogate}
\label{sec:eped}
\begin{figure}[htb]
    \centering
    \begin{subcaptionblock}{7cm}
        \centering
        \includegraphics[width=\linewidth]{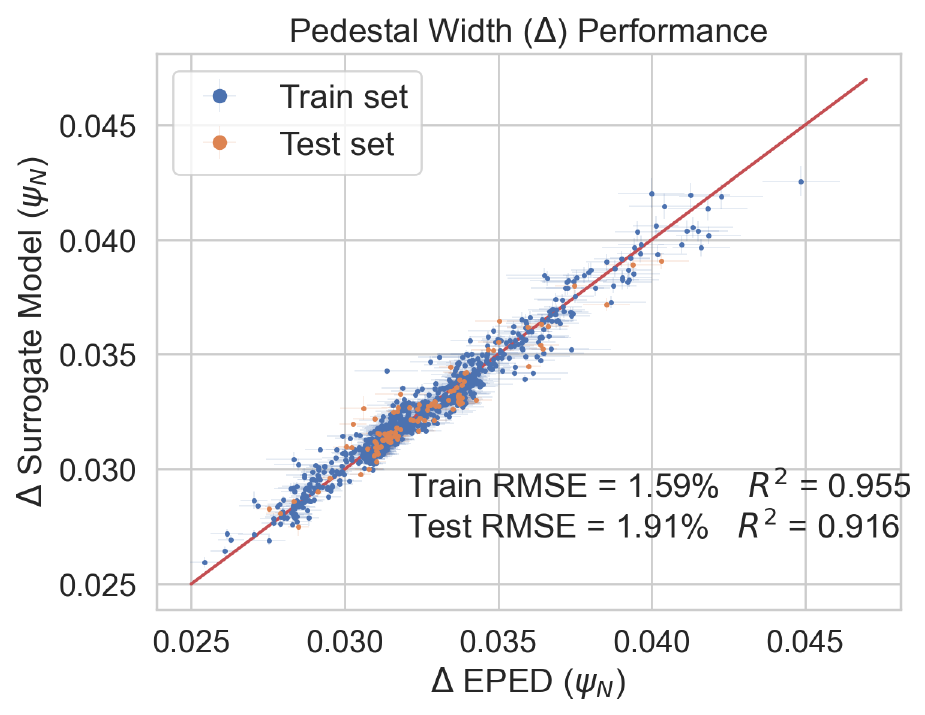}
        \caption{$\Delta$}
        \label{fig:EPED_perf_Delta}
    \end{subcaptionblock}%
    \vspace{0.1cm}
    \begin{subcaptionblock}{7cm}
        \centering
        \includegraphics[width=\linewidth]{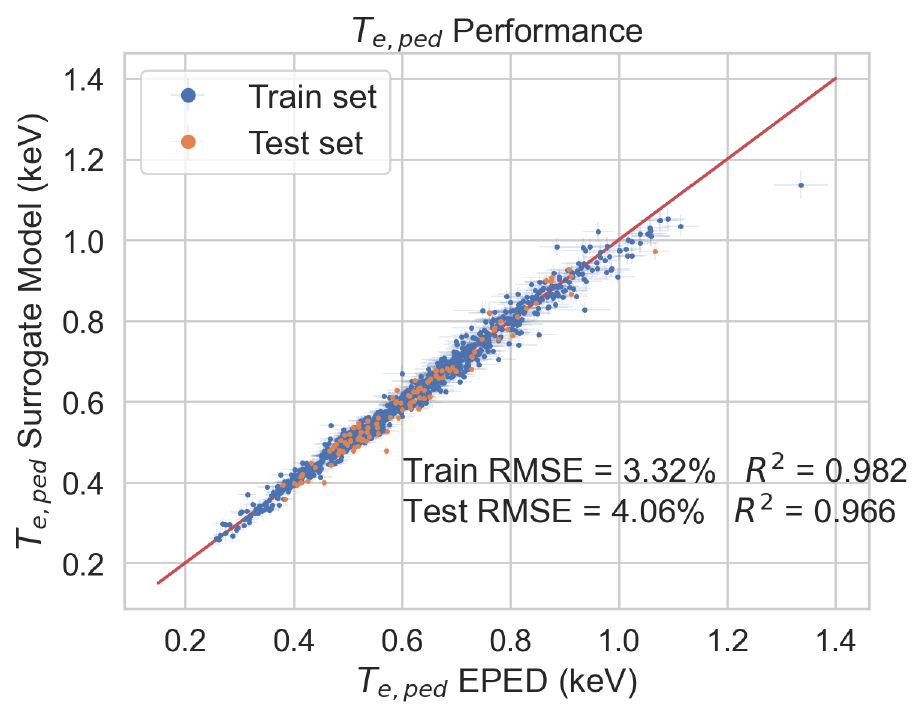}
        \caption{$T_{e,ped}$}
        \label{fig:EPED_perf_Te}
    \end{subcaptionblock}
    \caption{EPED surrogate performance for its two output variables. Train and test data are shown in blue and orange respectively. Epistemic and aleatoric uncertainties are displayed in vertical and horizontal axis respectively.}
    \label{fig:EPED_perf}
\end{figure}

Following the same principles and methodology developed for EuroPED-NN, and also to show the adaptability of it, the surrogate model of EPED has been created. Taking the inputs and outputs of EPED listed in the Section~\ref{subsec:ModelInputOutput}, a BNN-NCP with the same number of neurons per layer and the same architecture has been trained to check if the BNN captures the model and the $\Delta-\beta_{p,ped}$ relation. In this case the outputs will correspond to the values of $T_{e}$ at top pedestal ($T_{e,ped}$) and the pedestal width $\Delta$. The dataset used will be the same as in EuroPED-NN, where the columns for EPED runs were already present. The model has been trained for 1000 epochs. The performance is shown in the Figures~\ref{fig:EPED_perf_Delta} and \ref{fig:EPED_perf_Te}. Both of them seem to have a good fit with the 1:1 line, and the performance is always lower than 5 \% in both test and train datasets. RMSE is consistently lower than in EuroPED-NN, which evidence the lower complexity of EPED that allows better performance with the same neural network architecture.

\begin{figure*}[b]
    \centering
    \begin{subcaptionblock}{6cm}
        \centering
        \includegraphics[width=\linewidth]{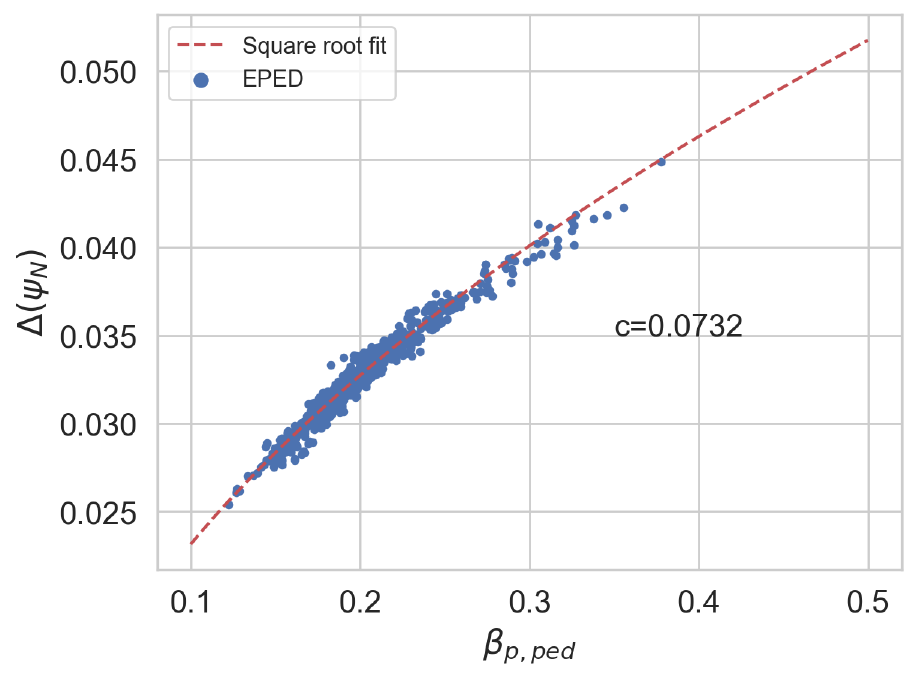}
        \caption{EPED dataset}
        \label{fig:beta_delta_eped_db}
    \end{subcaptionblock}%
    \begin{subcaptionblock}{6cm}
        \centering
        \includegraphics[width=\linewidth]{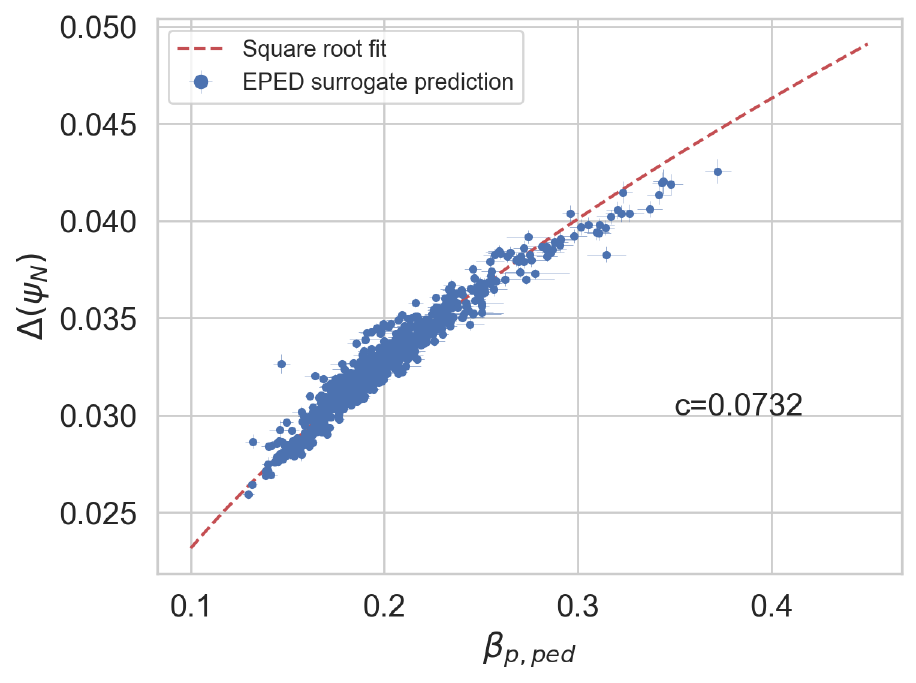}
        \caption{EPED surrogate predictions with dataset points}
        \label{fig:beta_delta_sur_eped}
    \end{subcaptionblock}%
    \begin{subcaptionblock}{6cm}
        \centering
        \includegraphics[width=\linewidth]{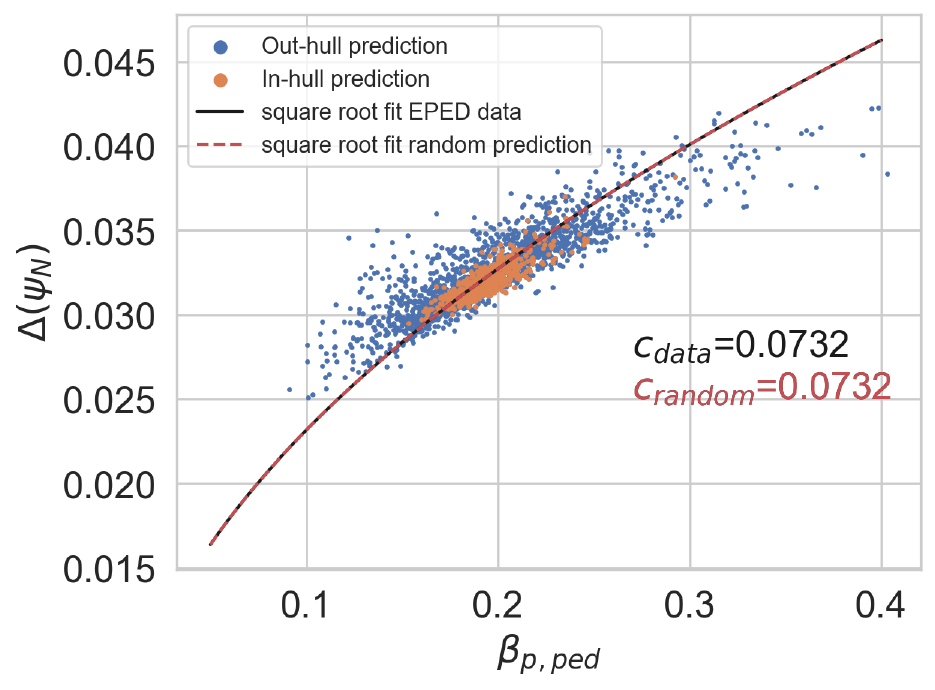}
        \caption{EPED surrogate predictions with random points}
        \label{fig:beta_delta_eped_rand}
    \end{subcaptionblock}
    \caption{Data from various sources represented in pedestal-width ($\Delta$)-$\beta_{p,ped}$ space, with errorbars showing the epistemic uncertainty. Predictions (blue dots) are fitted to show the semi empirical relation $\Delta=c\sqrt{\beta_{p,ped}}$ (square root fit in dashed red). The pedestal width is measured in normalised magnetic flux units $\psi_N$ and $\beta_{p,ped}$ is a dimensionless quantity. In \textbf{(c)}, prediction points inside the training convex hull are shown in orange.}
    \label{fig:beta_delta_eped}
\end{figure*}

As in EuroPED, the semi empirical relation $\Delta=c\sqrt{\beta_{p,ped}}$ is quite relevant in EPED, so we expect the EPED surrogate to hold the same relation. Then we will perform the physical validation as before. Figure~\ref{fig:beta_delta_eped_db} shows the EPED predictions and the square root relation in them. In Figure~\ref{fig:beta_delta_sur_eped} the predictions of the EPED surrogate are also shown in the pedestal width ($\Delta$)-$\beta_{p,ped}$ space, where the square root fit coefficient is almost the same as in the original EPED model. Finally in Figure~\ref{fig:beta_delta_eped_rand} 4000 random input entries were generated in the same manner as in Section~\ref{subsubsec:delta_beta} and they were passed through EPED surrogate. It is appreciable that prediction points that fall inside the training convex hull follow better the trend line, although the out-hull points also follow it reasonably well. So again, the surrogate model generalises correctly the physic relation.

Additionally, it makes sense to compare this EPED surrogate model with the model EPED1-NN \cite{aTGLFEPEDNN-Meneghini} that also replicates EPED using a neural network, as described in section \ref{sec:context}. In both cases the inputs are the same, although EPED1-NN is trained in tokamak DIII-D range of values and the EPED surrogate developed in this study is trained in JET range of values. Output variables are different only because EPED1-NN uses $p_{ped}$ and this study uses $T_{e,ped}$, however both of them are highly related. The architectures are different, EPED1-NN uses a regular feed-forward neural network while in this study the BNN-NCP technique has been applied to obtain uncertainty aware predictions. Regarding performance, EPED1-NN accuracy corresponds with $R^2=0.987$ and $R^2=0.964$ for $p_{ped}$ and pedestal width $\Delta$, respectively, while, as depicted in figure \ref{fig:EPED_perf}, EPED surrogate model accuracy lies around $R^2=0.966$ for $T_{e,ped}$ and $R^2=0.916$ for pedestal width $\Delta$. Both models have reasonably high accuracy for the intended outputs, although the one developed in this study is slightly lower.

Overall, in this section it has been shown the adaptability of the method to another plasma pedestal model, obtaining reasonably good performance and comparing it with a similar model.

\end{document}